\documentclass[review, 12pt, 3p, sort&compress]{elsarticle}

\usepackage{amssymb}

\usepackage{lineno}
\usepackage{wrapfig}
\usepackage{amsmath,amsthm,amsfonts}
\usepackage{url}
\usepackage{tikz}

\usepackage{subcaption}
\usepackage{todonotes}
\usepackage{lipsum}

\newcommand{\beginsupplement}{%
        \setcounter{table}{0}
        \renewcommand{\thetable}{S\arabic{table}}%
        \setcounter{figure}{0}
        \renewcommand{\thefigure}{S\arabic{figure}}%
		\setcounter{equation}{0}        
        \renewcommand{\theequation}{S\arabic{equation}}
        \setcounter{section}{0}
        \renewcommand{\thesection}{SM\arabic{section}}
}

\newcommand{\e}{\mathrm{e}}

\newcommand{\ud}{\mathrm{d}}

\newcommand{\erf}{\text{erf}}

\renewcommand{\vec}[1]{\mathbf{#1}}

\newcommand{\be}{\begin{equation}}
\newcommand{\ee}{\end{equation}}
\newcommand{\bea}{\begin{eqnarray}}
\newcommand{\eea}{\end{eqnarray}}
\newcommand{\beas}{\begin{eqnarray*}}
\newcommand{\eeas}{\end{eqnarray*}}
\newcommand{\bi}{\begin{itemize}}
\newcommand{\ei}{\end{itemize}}

\newcommand\nc{\newcommand}
\nc{\Oc}{\mathcal{O}} \nc{\Omo}{\Omega_{\tiny{\mbox{out}}}}
\nc{\Omi}{\Omega_{\tiny{\mbox{in}}}}
\nc{\Omit}{\Omega_{\tiny{\mbox{int}}}}
\nc{\Pio}{\Pi_{\tiny{\mbox{out}}}}
\nc{\Pii}{\Pi_{\tiny{\mbox{in}}}}
\nc{\Piit}{\Pi_{\tiny{\mbox{int}}}}
\nc\pa{\partial}
\nc\pad[2]{\frac{\pa #1}{\pa #2}} 
\nc\padd[2]{\frac{\pa^2 #1}{\pa{#2}^2}}
\nc\nd[2]{\frac{\ud #1}{\ud #2}}
\nc\ndd[2]{\frac{d^2 #1}{d {#2}^2}}
\nc\pat[2]{\frac{D #1}{D#2}} 
\nc\ov{\overline} 
\nc\degree{^{\circ}} 
\nc\ord[1]{{\cal O}(#1)} 
\nc\ra{\rightarrow} 
\nc\Ra{\Rightarrow} 
\nc\dint{{\mbox ~d}}
\nc\dg{{\dot \mu}}
\nc\units[1]{$^\text{#1}$}

\usepackage[normalem]{ulem}
\nc \mhtodo[1]{\todo[inline, backgroundcolor=blue!30]{#1}}
\nc \mh[1]{\textcolor{blue}{#1}}
\nc \mhso[1]{\textcolor{blue}{\sout{#1}}}
\nc \mc[1]{\textcolor{red}{#1}}
\definecolor{cgreen}{rgb}{0.0, 0.42, 0.24}
\nc \tim[1]{\textcolor{cgreen}{#1}}

\begin{document}
\begin{frontmatter}

\title{The One-Dimensional Stefan Problem with Non-Fourier Heat Conduction}%

\author[Tim,Tim2]{Marc Calvo-Schwarzw\"alder\corref{cor1}}
\author[Tim,Tim2]{Timothy~G. Myers}
\author[Matt]{Matthew~G. Hennessy}

\cortext[cor1]{mcalvo@crm.cat}
\address[Tim]{Centre de Recerca Matem\`{a}tica, Campus de Bellaterra, Edifici C, 08193 Bellaterra, Barcelona, Spain.}
\address[Tim2]{Departament de Matem\`{a}tiques, Universitat Polit\`{e}cnica de Catalunya, 08028 Barcelona, Spain.}
\address[Matt]{Mathematical Institute, University of Oxford, Oxford, OX2 6GG, United Kingdom.}

\begin{abstract}
	We investigate the one-dimensional growth of a solid into a liquid bath, starting from a small crystal, using the Guyer-Krumhansl and Maxwell-Cattaneo models of heat conduction. By breaking the solidification process into the relevant time regimes we are able to reduce the problem to a system of two coupled ordinary differential equations describing the evolution of the solid-liquid interface and the heat flux. The reduced formulation is in good agreement with numerical simulations. In the case of silicon, differences between classical and non-classical solidification kinetics are relatively small, but larger deviations can be observed in the evolution in time of the heat flux through the growing solid. From this study we conclude that the heat flux provides more information about the presence of non-classical modes of heat transport during phase-change processes.
\end{abstract}


\end{frontmatter}



\section{Introduction}\label{sec:intro}
Heat conduction at small length or time scales is dominated by effects which are not well described by the classical equations \cite{Cahill2003}. For instance, it has been observed that the thermal conductivity, which was introduced as a material-intrinsic property and therefore independent of size or shape, exhibits size-dependent behaviour at small length scales \cite{Li2003}. The understanding and correct description of the underlying thermal physics is of vital importance for future applications of nanostructures. For instance, an incorrect description of the thermal response of nanoscale devices can lead to melting and eventually device failure \cite{Nie2011}. Novel applications such as laser melting involve phase change at very short time scales \cite{Glicksman2010,Asta2009,Liu2006}. 

When the characteristic length or time scales of a physical system are comparable to the characteristic length or time scales of the heat carriers, non-classical effects are expected to dominate the heat-transfer mechanisms \cite{Cahill2003,Siemens2010,Torres2018,Calvo2018,Calvo2018b}. In semiconductors such as silicon or germanium, these energy carriers are predominantly phonons, which may be understood as lattice vibrations. These quasi-particles have two associated important physical quantities: the mean free path and the thermal relaxation time. From a microscopic point of view, these represent the mean distance and time that phonons travel between collisions. We can then distinguish two thermal relaxation times, normal and resistive, depending on whether or not phonon momentum is conserved after the collision. 

A number of different models extending the classical Fourier law can be found in the literature. In this paper we focus our attention to the Maxwell-Cattaneo equation \cite{Cattaneo1958,Vernotte1958} and the Guyer-Krumhansl equation \cite{Guyer1966a,Guyer1966b}. Other popular continuum models are the dual phase lag model \cite{Tzou1995,Tzou1996} or the thermomass model \cite{Wang2010a,Cao2007,Guo2010}. The first incorporates a time lag into all the terms appearing in Fourier's law, although the physical interpretation of the new parameters remains unclear. The latter uses the Einstein energy-mass relation to describe heat conduction as the motion of a gas formed by particles with non-zero mass called thermons. In this case, a non-linear term emerges in the governing equations, which complicates the mathematical analysis. In contrast to these models, the Guyer-Krumhansl and Maxwell-Cattaneo models involve linear equations where the parameters have a clear physical meaning.

The Maxwell-Cattaneo equation was proposed in the 1950s to remove the infinite speed of heat propagation implicit to Fourier's law. Mathematically, this model extends Fourier's law by incorporating a new term which accounts for the time lag between the imposition of a temperature gradient and the creation of a heat flux. This new term leads to the flux depending on the history of the temperature gradient and is thus referred to as a memory term. In the 1960s Guyer and Krumhansl \cite{Guyer1966a,Guyer1966b} proposed an extension to Fourier's law derived from the Boltzmann transport equation (BTE) \cite{Boltzmann1872} which accounts for memory and non-local effects. Whereas memory effects are important for short times or high frequencies, non-local effects become relevant at small length scales. Although this formalism was initially thought to be applicable only in low-temperature situations ($\sim 2$ K), recent experimental observations show that it is also suitable for describing heat transfer at room temperature or above \cite{Torres2018,Guo2018,Lee2015}. Jou et al.~\cite{Jou1996} later showed that both the Maxwell-Cattaneo and the Guyer-Krumhansl equations can be derived using the framework of extended irreversible thermodynamics (EIT). 

The classical mathematical description of a phase-change process, known as the Stefan problem, is based on Fourier's law \cite{Gupta2003}. Some authors have incorporated the Maxwell-Cattaneo equation into the formulation of the Stefan problem to model rapid solidification processes \cite{Sobolev2015}, cryopreservation \cite{Deng2003,Ahmadikia2012}, cryosurgery for lung cancer treatments \cite{Kumar2017}, and nanoparticle melting \cite{Hennessy2019}.
Other authors have studied the Stefan problem with Maxwell-Cattaneo conduction from a point of view of mathematical analysis \cite{Glass1991,Greenberg1987,Solomon1985}. Recently, Hennessy et al.~\cite{Hennessy2018} performed a detailed asymptotic analysis of the one-dimensional Stefan problem with Guyer-Krumhansl conduction, where it is shown that non-classical effects can lead to important differences in the solidification kinetics with respect to Fourier's law. Sobolev \cite{Sobolev1996,Sobolev1991} solved different non-Fourier formulations under the assumption of constant interface velocity. To account for non-local effects, some authors \cite{Font2018,Calvo2019a,Calvo2019b} recently proposed a formulation of the Stefan problem based on Fourier's law with an effective thermal conductivity dependent on the size of the solid. From a microscopic point of view, Font and Bresme \cite{Font2018b} published a work where non-classical effects in a melting process were observed using molecular dynamics. 

In this paper, we aim to provide tools for detecting non-classical heat transport mechanisms in phase-change processes. For this we study the solidification kinetics in a one-dimensional geometry using the Maxwell-Cattaneo and Guyer-Krumhansl models of heat conduction. The solid is assumed to grow from a small seed crystal, which splits the process into two major regimes, with the first  capturing the heat conduction through the seed crystal and the second describing the solidification kinetics. Our asymptotic analysis shows that these two time regimes can be split into multiple sub-regimes, with each of them describing different physical phenomena. 
We show that, although both conduction models describe the initial transport of heat through the seed crystal in different ways, the solidification process can be captured by reducing the full problem to the same system of ordinary differential equations, which provides excellent agreement with numerical simulations. The results are then applied to the solidification of silicon, where we conclude that the motion of the interface does not necessarily give information about non-Fourier heat transport. Conversely, the evolution of the heat flux is able to provide valuable insights about non-classical modes of heat transfer.

\section{Mathematical formulation}\label{sec:model}
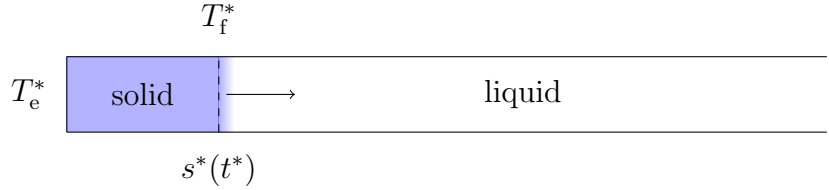
\begin{figure}[h!]
\centering
	\begin{tikzpicture}
	\fill[blue,opacity=0.3] (0,0) rectangle (2,1);
	\shade[left color=blue,right color=blue!.5!white,opacity=0.3] (2,0) rectangle (2.2,1);
	\draw (0,0) -- (10,0);
	\draw (0,1) -- (10,1);
	\draw (0,0) -- (0,1);
	\draw [dashed](2,0) -- (2,1);
	\draw [->] (2.1,.5) -- (3,.5);
	\node at (1,.5) {solid};
	\node at (6,.5) {liquid};
	\node at (2,-.5) {$s^*(t^*)$};
	\node at (2,1.5) {$T^*_\text f$};
	\node at (-.5,.5) {$T^*_\text e$};
	\end{tikzpicture}
	\caption{A semi-infinite bath, initially at the phase change temperature $T^*_\text{f}$, that is solidifying from $x^*=0$ due to a low temperature in the surrounding environment $T^*_\text{e}<T^*_\text{f}$. The solid-liquid interface is located at $x^*=s^*(t^*)$.}\label{fig:bar}
\end{figure}

To elucidate the roles of non-Fourier conduction mechanisms during solidification, we consider a simple one-dimensional geometry consisting of a liquid bath, which is initially at the phase change temperature $T^*_\text{f}$, where the $^*$ notation denotes dimensional quantities. Due to a cold temperature $T^*_\text{e}$ in the environment in contact with the bath at $x^*=0$, the liquid starts solidifying and a solid phase grows into the bath, occupying the space $[0,s^*(t^*)]$. A schematic of the physical scenario is depicted in Fig.~\ref{fig:bar}. By assuming that the liquid phase is at the phase-change temperature, any decrease in temperature will cause solidification to occur. In this way we can focus on the heat transfer in the solid, which drives the solidification process and is where the non-classical phenomena are expected to occur. 

\subsection{Governing equations}
Conservation of energy in the solid is described by
\begin{equation}\label{cons_energy}
	c^*\rho^*\pad{T^*}{t^*}+\pad{Q^*}{x^*}=0,
\end{equation}
where  $Q^*$, $T^*$, $c^*$ and $\rho^*$ are respectively the heat flux, the temperature, the specific heat capacity and the density of the solid. In the classical formulation, the mathematical description is completed by assuming that the heat flux is determined by Fourier's law, which in one dimension is
\begin{equation}\label{Fourier}
    Q^*=-k^*\pad{T^*}{x^*},
\end{equation}
where $k^*$ is the bulk thermal conductivity of the solid. In the present study this will be replaced by the Maxwell-Cattaneo equation (MCE) or the Guyer-Krumhansl (GKE). In one spatial dimension, the GKE takes the form
\begin{equation}\label{GK:Eq}
	\tau^*_R\pad{Q^*}{t^*}+Q^*=-k^*\pad{T^*}{x^*}+3\ell^{*2}\pad{^2Q^*}{x^{*2}},
\end{equation}
where $\tau^*_R$ is the resistive thermal relaxation time, which is the mean time between consecutive phonon collisions which do not conserve momentum, and $\ell^*$ represents the phonon mean free path, i.e. the mean distance between two collisions. The factor 3 in the last term appears due to different second-order terms collapsing to a single contribution in one dimension. The first term of the left-hand side and the second term of the right-hand side of \eqref{GK:Eq} represent respectively memory and non-local effects. The prior are important at small time scales and therefore critical for the initial stages of the solidification process. 
Non-local effects appear due to two reasons. Firstly, they are caused by the non-negligible rate of normal scattering, i.e. phonon collisions are more likely to conserve their momentum and relaxation effects are delayed. Secondly, non-local effects are important when the characteristic size of the system is comparable to the mean free path.

The GKE can be reduced to the MCE by neglecting the term corresponding to non-local effects, i.e. by setting $\ell^*=0$ in Eq.~\eqref{GK:Eq}. The two approaches, although similar, are based on different frameworks:  the GKE is derived from the BTE whereas the MCE is a phenomenological approach to account for finite heat carrier speed. However, it was later shown that both can be derived within the EIT framework \cite{Jou1996,Lebon2014}.

\subsection{Boundary and initial conditions}
At the boundary $x^*=0$ we impose a fixed-temperature condition given by
\begin{equation}\label{model:bc0}
  T^*=T^*_\text{e},\qquad \text{at }x^*=0^*.
\end{equation}
A boundary condition of the form of Eq.~\eqref{model:bc0} predicts an initially infinite rate of phase change \cite{Gupta2003,Alexiades1992,Font2013,Font2015} when the growing phase initially has zero size. In practice, other forms of boundary conditions should be applied, such as the Newton cooling condition. This has been used in the modelling of melting of nanoparticles \cite{Ribera2016} and nanowires \cite{Florio2016}, and leads to a significant increase of the melting times. The cooling condition has also been used in a recent study by Hennessy et al. \cite{Hennessy2018}, where a thorough asymptotic analysis of the Guyer-Krumhansl-Stefan problem is performed. The same authors also considered a cooling condition accounting for memory effects when studying the melting behaviour of nanoparticles using the MCE \cite{Hennessy2019}. In this problem, we will assume that solidification begins from a seed crystal of finite size and thus the condition \eqref{model:bc0} can be used without inducing an unphysical infinite rate of phase change.

At the moving interface we impose that the temperature is equal to the phase-change temperature,
\begin{equation}\label{model:bcs}
	T^*=T^*_\text{f},\qquad \text{at }x^*=s^*,
\end{equation}
which implies continuity in temperature across the interface. Other authors claim that a jump condition for the temperaure should be utilized at the interface \cite{Glass1991,Greenberg1987,Sobolev1995}. Based on a diffuse interface model, Hennessy et al.~\cite{Hennessy2019} show that, when the MCE is used to model the melting of nanoparticles, the jump condition avoids the phenomenon of supersonic melting \cite{Sobolev1991,Sobolev2015,Mullis1997}, which may appear when \eqref{model:bcs} is used. In the case of nanoparticle melting, the onset of supersonic phase change is due to the spherical geometry of system and the fact that less energy is required to melt the surface of smaller solid cores. Thus, the rate of melting increases and eventually becomes unbounded as the radius of the nanoparticle decreases to zero. As there is no mechanism for supersonic solidification in this one-dimensional Cartesian problem, it is sufficient to consider the case of temperature continuity, which is derived from the jump condition in the limit of slow phase change.

In its simplest form, the Stefan condition is given by
\begin{equation}\label{model:Stefan}
	\rho^* L^*_m \nd{s^*}{t^*}=-Q^*,\qquad \text{at }x^*=s^*,
\end{equation}
where $L^*_m$ is the latent heat.  We assume that solidification starts from an initial seed crystal of size $s_c^*$, formed, for instance, by heterogeneous nucleation, where the temperature is equal to the temperature of the liquid. Since we are using continuum models to describe the growth of the solid, additional assumptions on the size of the seed crystal need to be made. We will assume that its thickness is of the order of nanometres, large enough for continuum theory to hold \cite{Myers2014}. The initial conditions are therefore
\begin{equation}\label{model:ic}
	s=s_c^*,\qquad T^*=T^*_\text{f},\qquad Q^*=0,\qquad \text{at }t^*=0.
\end{equation}

\subsection{Alternative formulations}
Conservation of energy \eqref{cons_energy} and the GKE \eqref{GK:Eq} can be combined to give
\begin{equation}\label{GK2}
	\tau^*_R\pad{Q^*}{t^*}+Q^*=-k^*\left(\pad{T^*}{x^*}+\frac{3\ell^{*2}}{\alpha^*}\pad{^2T^*}{x^*\partial t^*}\right),
\end{equation}
with $\alpha^*=k^*/(\rho^*c^*)$ being the bulk thermal diffusivity. This form shows that non-local effects are related to instantaneous changes of the temperature gradient, whereas memory effects are described temporal changes of the flux (which can be related to the time integral of the temperature gradient). In the special case where $\tau^*_R=3\ell^{*2}/\alpha^*$, it is easy to see that Fourier's law is recovered from \eqref{GK2}. This phenomenon has been termed Fourier resonance \cite{Both2016,Van2017}.  Alternatively, it is possible to eliminate the flux from the equation to obtain the Guyer-Krumhansl heat equation (GKHE)
\begin{equation}\label{GKHE}
	\tau^*_R\padd{T^*}{t^*}+\pad{T^*}{t^*}=\alpha^*\padd{T^*}{x^*}+3\ell^{*2}\pad{^3T^*}{x^{*2}\partial t^*}.
\end{equation}
In the limit $\ell^*\to0$ we obtain the heat equation for the MCE, which in the literature is termed the hyperbolic heat equation (HHE).

\subsection{Non-dimensional formulation}\label{ssec:nd}
Since the classical solidification kinetics are expected to be recovered for sufficiently large times, we will employ the typical scales used to non-dimensionalise the classical equations. In this way, the importance of non-classical contributions will be described by new dimensionless parameters.

The natural length scale of the problem is the initial size of the solid, $x^*,s^*=O( s^*_c)$. Time is then scaled as $t^*=O(\tau^*_D)$ with $\tau^*_D=s^{*2}_c/\alpha^*$, which corresponds to the time scale for diffusion through the initial seed crystal. The temperature scale is defined by the difference between the freezing and environment temperatures, $\Delta T=T^*_\text{f}-T^*_\text{e}$. Balancing terms in \eqref{cons_energy} then requires $Q^*=O(Q^*_0)$, where $Q^*_0=k^*\Delta T/s^*_c$. Based on these typical scales we introduce the following non-dimensional variables:
\begin{equation}\label{model:nd_vars}
	x=\frac{x^*}{s^*_c},\qquad t = \frac{t^*}{\tau^*_D},\qquad T=\frac{T^*-T^*_\text{f}}{\Delta T},\qquad Q= \frac{Q^*}{Q^*_0},\qquad s=\frac{s^*}{s^*_c}.
\end{equation}
In these variables, conservation of energy \eqref{cons_energy} and the GKE \eqref{GK:Eq} become
\begin{subequations}\label{nd:eqs}
  \begin{align}
    T_t +Q_x&=0, \label{nd:ce} \\
    \gamma Q_t +Q &=-T_x +\eta^2Q_{xx},\label{nd:GK}
  \end{align}
\end{subequations}
where we have introduced the subscript notation to denote derivatives. The new parameters in \eqref{nd:GK} are $\gamma=\tau^*_R/\tau^*_D$ and $\eta=3^{1/2}\ell^*/s_c^*$, which represent the dimensionless forms of the relaxation time and mean free path. In the literature, $\gamma$ is often called the Cattaneo number, whereas $\eta$ is proportional to the ratio of the mean free path to the initial size of the system and termed the Knudsen number. 

The non-dimensional boundary, Stefan, and initial conditions become
\begin{subequations}\label{nd:bc}
\begin{alignat}{3}
	&T=-1,\qquad &\text{at }&x=0,\label{nd:bc0}\\
	&T=0,\qquad &\text{at }&x=s(t),\label{nd:bcs}\\
	&\beta s_t=-Q,\qquad &\text{at }&x=s(t),\label{nd:Stefan}\\
	&T=Q=0,\quad s=1,\qquad &\text{at }&t=0,\label{nd:ic}
\end{alignat}
\end{subequations}
where $\beta=L_m^*/(c^*\Delta T)$ is the Stefan number and represents the ratio of latent heat to sensible heat. 

In the dimensionless formulation, \eqref{GK2} becomes
\begin{equation}\label{nd:GK2}
	\gamma Q_t+Q=-T_x-\eta^2T_{xt},
\end{equation}
and Fourier resonance appears if the non-classical dimensionless parameters coincide, $\gamma=\eta^2$. Finally, the dimensionless GKHE \eqref{GKHE} then takes the form
\begin{equation}\label{nd:GKHE}
	\gamma T_{tt}+T_t=T_{xx}+\eta^2T_{xxt}.
\end{equation}

\subsection{Parameter estimation}
\label{ssec:parameterestimation}
Silicon is a semiconductor of high interest for many nanoscale applications. The phase-change temperature and latent heat of silicon are $T^*_\text{f}=1687$ K and $L^*_\text{m}=1787\times10^3$ J/kg \cite{Mills2000}. Since the coefficient of thermal expansion is of the order of $10^{-6}$ K$^{-1}$ \cite{Okada1984}, we can assume a constant value for the density. Based on Mills and Courtney \cite{Mills2000}, we take $\rho^*=2296$ kg/m$^3$. Near the phase-change temperature, the thermal conductivity and specific heat capacity of silicon are $k^*=22.1$ W/m$\cdot$K and $c^*=1032$ J/kg$\cdot$K \cite{EfundaSilicon,Hull1999}. The thermal diffusivity is therefore $\alpha^*=9.33\times10^{-6}$ m\units{2}/s.

\begin{figure}
    \centering
    \includegraphics[width=.47\textwidth]{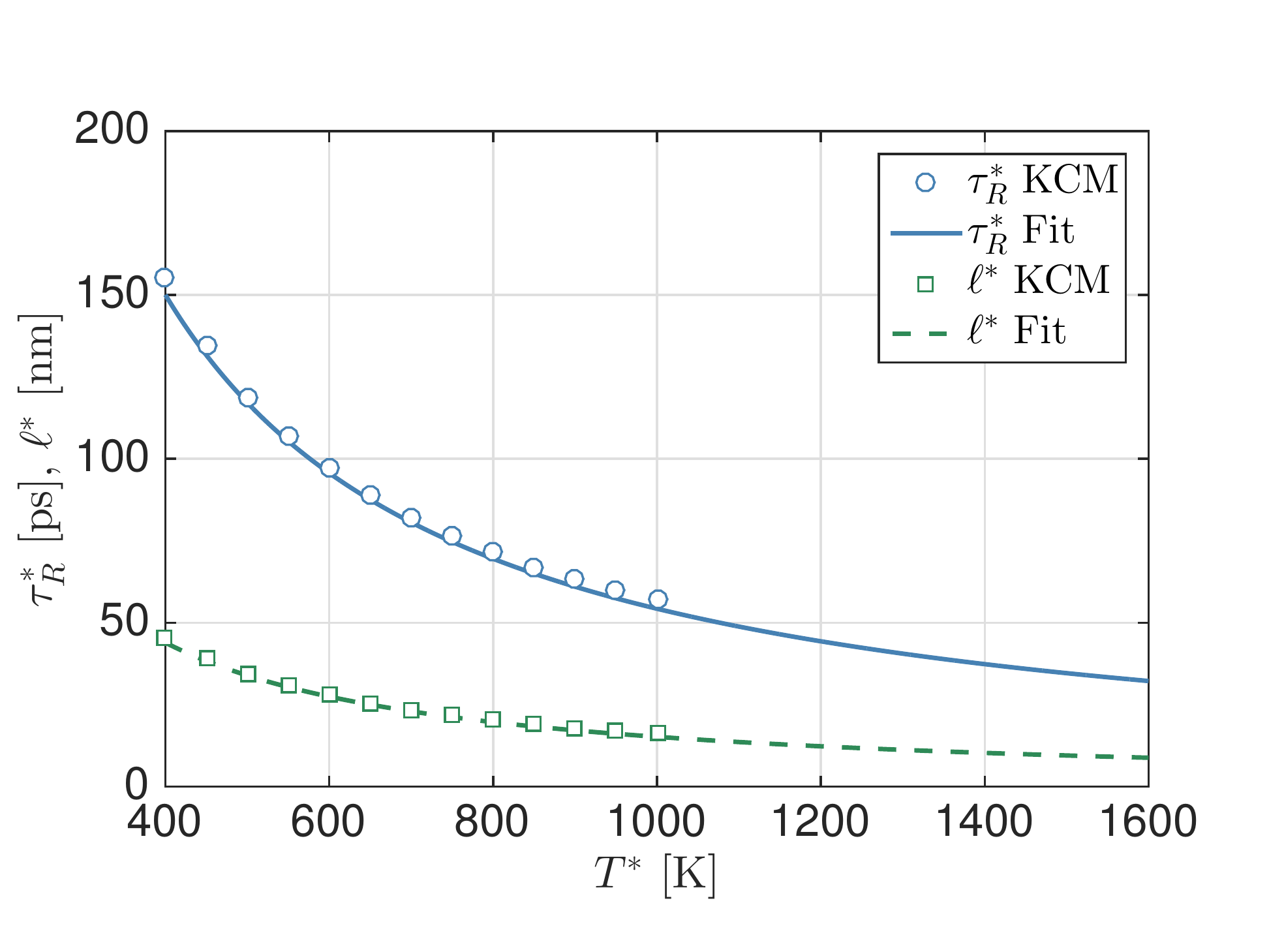}
	\caption{Relaxation time and mean free path of phonons for silicon in the range $400$ K$<T^*<1600$ K. Symbols refer to data provided by the KCM \cite{Torres2017b}, whereas lines correspond fitting those values to an expression of the form $a\cdot(T^*)^{-b}$.} \label{fig:propertiesKCM}
\end{figure}

Determining the relaxation time $\tau^*_R$ and the mean free path $\ell^*$ is a more difficult task. Here we use the kinetic collective model (KCM) \cite{De2014}, which has been demonstrated to be a powerful tool for predicting the thermal properties of various materials \cite{Torres2017b,Torres2018}. Since the calculation time increases and the accuracy decreases as we increase the temperature, in Fig.~\ref{fig:propertiesKCM} we show results of the KCM up to 1000 K and then fit these values with an equation of the form $a\cdot(T^*)^{-b}$ to find approximate values of the non-classical parameters near $T^*_\text f$. We find $a\approx1.16\times10^{-7}$ and $b\approx1.11$ for the relaxation time, whereas for the mean free path the best fit is obtained taking $a\approx4.58\times10^{-5}$ and $b\approx1.16$. From these expressions we find $\tau^*_R\approx32.16$ ps and $\ell^*\approx8.79$ nm at $T^*=1600$ K.

\begin{table}[]
    \centering
    \caption{Values of the thermophysical properties of silicon used in this study \cite{Mills2000,Torres2017b,EfundaSilicon,Hull1999}.}
    \label{tab:silicon}
    \begin{tabular}{c c c c}
         \hline
         Property & Symbol & Value & SI Units  \\
         \hline
         \hline
         Thermal conductivity & $k^*$ & 22.1 & W/m$\cdot$K \\
         Specific heat capacity & $c^*$ & 1032 & J/kg$\cdot$K \\
         Density & $\rho^*$ & 2296 & kg/m\units{3}\\
         Thermal diffusivity & $\alpha^*$ & 9.33$\times$10\units{-6} & m\units{2}/s\\
         Phase change temperature & $T^*_\text{f}$ & 1687 & K\\
         Latent heat & $L^*_\text{m}$ & 1787$\times$10\units{3} & J/kg\\
         Relaxation time & $\tau^*_R$ & 32.16 & ps\\ 
         Mean free path & $\ell^*$ & 8.79 & nm\\ 
         \hline
    \end{tabular}
\end{table}

Independent of the underlying model, the dynamics of the solidification process are affected by the Stefan number $\beta$, which can be parametrized in terms of the temperature scale as $\beta=\mathcal{T}^*/\Delta T$. For silicon we find $\mathcal{T}^*\approx1731.6$ K, hence even by setting the environmental temperature to $0$ K (thus $\Delta T=1687$ K) we find $\beta>1$. Since our study involves an external temperature that is far above absolute zero, we therefore assume $\beta\gg1$, in accordance with previous studies.

The Cattaneo number $\gamma$ represents the ratio of the time scales of the ballistic and diffusive heat transport regimes. For $\gamma\ll1$, heat transfer is dominated by diffusion, whereas it becomes ballistic for $\gamma\gg1$. The Cattaneo number may be parametrized as $\gamma=(L^*/s_c^*)^2$, where $L^*=\sqrt{\alpha^*\tau^*_R}$ is an effective length where memory effects become non-negligible. Upon using the values given in Table~\ref{tab:silicon}, we find $L^*\approx17.32$ nm. The Knudsen number $\eta$ corresponds to the ratio of the size of the initial crystal seed to the mean free path and therefore characterises the impact of non-local effects on the initial transport of heat. Using the values in Table~\ref{tab:silicon} we obtain $\eta\approx\ell^*_\text{eff}/s_c^*$, where $\ell^*_\text{eff}=3^{1/2}\ell^*\approx15.23$ nm is the effective mean free path.  Since we assume the size of the seed crystal to be of the order of nanometres, we expect $\gamma,\eta\gg1$ and thus heat conduction through the solid will initially be dominated by non-classical effects.

Finally, the typical scale of the flux may be parametrized in terms of the Stefan number and the initial size of the seed crystal, $Q_0^*=(k^*L^*_m/c^*)\times(\beta s_c^*)^{-1}$. For silicon, $k^*L_m^*/c^*\approx 3.8\times 10^{-4}$~W/m.

\section{Numerical solution}\label{sec:numerics}
To solve the problem numerically, we transform the growing domain $[0,s(t)]$ to the unit interval $[0,1]$ and then use finite differences to reformulate the problem as a set of algebraic equations.

\subsection{Formulation in a fixed domain}
Upon introducing the alternative space coordinate $\xi=x/s(t)$, which transforms the growing domain into the unit interval, the derivatives become
\begin{equation}
    \pad{}{t}\mapsto\pad{}{t}-\xi\frac{s_t}{s}\pad{}{\xi},
    \qquad \pad{}{x}\mapsto\frac{1}{s}\pad{}{\xi},
    \qquad \padd{}{x}\mapsto\frac{1}{s^2}\padd{}{\xi}. 
\end{equation}
To avoid performing this transformation on the second-order time derivative in \eqref{nd:GKHE}, which becomes a particularly complicated expression, we introduce an auxiliary variable $v = T_t$ to write the problem solely in terms of first-order time derivatives. Hence, we let $T(x,t)=u(\xi,t)$, $T_t(x,t)=v(\xi,t)$ and $Q(x,t)=w(\xi,t)$. In these variables, $u$ and $v$ are determined by
\begin{subequations}\label{num:system1}
\begin{align}
        \gamma v_t -\gamma\frac{s_t}{s}\xi v_\xi + v -\eta^2\frac{1}{s^2}v_{\xi\xi} -\frac{1}{s^2}u_{\xi\xi} &=0,\\
        v-u_t+\frac{s_t}{s}\xi u_\xi&=0,
\end{align}
which are obtained from the GKHE \eqref{nd:GKHE} and the definition of $v$. Evaluating \eqref{nd:ce} at $t=0$ and applying the initial condition $Q=0$ (hence $Q_x=0$ at $t=0$) yields $T_t=0$ at $t=0$. From the boundary conditions \eqref{nd:bc0}-\eqref{nd:bcs} we find
\begin{alignat}{3}
        &u=-1,\qquad &\text{at }&\xi=0,\\
        &u=0,\qquad &\text{at }&\xi=1,\\
        &v=0,\qquad &\text{at }&\xi=0,\\
        &v+\frac{s_t}{s}u_\xi=0,\qquad &\text{at }&\xi=1.\label{num:bc_v_1}
\end{alignat}
\end{subequations}
The transformed flux $w$ can be determined by the transformed GKE
\begin{subequations}\label{num:system2}
\begin{align}
    &\gamma w_t-\gamma\frac{s_t}{s}\xi w_\xi+w- \eta^2\frac{1}{s^2} w_{\xi\xi}=-\frac{1}{s}u_{\xi}.\label{num:Q_fixed}
\end{align}
Assuming that conservation of energy holds at $\xi=0,1$, we find the boundary and initial conditions 
\begin{alignat}{5}
        w_\xi&=0,\qquad &\text{at }&\xi=0,\\
        w_\xi&=s_tu_\xi,\qquad &\text{at }&\xi=1,\\
        w&=0,\qquad &\text{at }&t=0.
\end{alignat}
\end{subequations}
Note, in principle no boundary conditions for $w$ are needed since we could use \eqref{nd:GK2} to write \eqref{num:Q_fixed} in terms of $w$, $w_t$, and derivatives of $u$ and $v$, the latter of which are already known from \eqref{num:system1}.

Finally, the transformed Stefan condition is
\begin{equation}\label{num:Stefan}
	\beta s_t=-w,\qquad \text{at }\xi=1.
\end{equation}

From this formulation we obtain the numerical solution by solving \eqref{num:system1}, \eqref{num:system2} and \eqref{num:Stefan} at each time step using finite differences. Further details about the numerical scheme are provided in the Supplementary Material.

\subsection{Small-time approximations}
Numerical experimentation has shown that replacing the initial conditions for $u$, $v$, and $w$ with approximate solutions that are valid for arbitrarily small times $t \ll 1$ greatly improves the stability of the computational scheme when solving the GKE (for $\eta \neq 0$).  By carrying out a small-time analysis of the model, which is described in the Supplementary Material, we find that the solution at time $t_0 \ll 1$ can be approximated by 
\begin{subequations}
\begin{align}
    &u\approx-1+\erf\left(\frac{\xi}{2\sqrt{\zeta t_0}}\right),\\
    &v\approx-\frac{\xi}{2\sqrt{\pi\zeta t_0^3}}\exp\left(-\frac{\xi^2}{4\zeta t_0}\right),\\
    &w\approx-\sqrt{\frac{\zeta}{\pi t_0}}\exp\left(-\frac{\xi^2}{4\zeta t_0}\right).
\end{align}
\end{subequations}
where $\zeta=\eta^2/\gamma$ represents an effective thermal conductivity.  When solving the MCE, the original initial conditions $u(\xi,0)=v(\xi,0)=w(\xi,0)=0$ can be used.

\section{Reduction of the equations}\label{sec:reduction}
As shown in the asymptotic analysis in the Supplementary Material, by the time that solidification occurs, the heat flux becomes approximately uniform in space: $Q_x \simeq 0$. This leads to the so-called quasi-steady regime, since differentiating the GKE \eqref{nd:GK} with respect to $x$ then leads to $T_{xx}\simeq0$ and hence time is present only due to the boundary condition \eqref{nd:bcs}. This quasi-steady regime also appears in the classical formulation under the assumption $\beta\gg1$, allowing the system to be reduced to a simple ordinary differential equation (ODE) for the position of the interface \cite{Davis2001}. Our asymptotic analysis reveals that a similar procedure can be considered here even in the case of strongly non-classical effects. Applying the boundary conditions \eqref{nd:bc0} and \eqref{nd:bcs} implies that the temperature is given by
\begin{equation}
  T(x,t)=\frac{x}{s}-1.
  \label{eqn:quasi_T}
\end{equation}
Upon using \eqref{eqn:quasi_T} in \eqref{nd:GK2}, and then \eqref{nd:Stefan} to eliminate the arising term with $s_t$, yields
\begin{equation}\label{reduction:Q_ode}
    Q_t=-\gamma^{-1}\left[\frac{1}{s}+\left(\frac{\eta^2}{\beta s^2}+1\right)Q\right].
\end{equation}
The asymptotic analysis, described in the Supplementary Material, shows that the appropiate matching condition is given by
\begin{equation}\label{reduction:Q_ic}
    Q(0)=-\eta^2/\gamma.
\end{equation}
Note, this initial condition is valid for both the Guyer-Krumhansl and Maxwell-Cattaneo models, since the limit $\eta\to0$ gives the initial condition $Q(0)=0$ which corresponds to the small time behaviour of the MC-Stefan problem. Hence, based on the asymptotic analysis we are able to reduce a problem involving two partial differential equations in a growing domain into a pair of coupled ordinary differential equations \eqref{nd:Stefan} and \eqref{reduction:Q_ode}.

\section{Results and discussion}\label{sec:res}
We are now in position to compare the different solidification kinetics in the limits of dominant non-classical effects and to assess the validity of the reduced formulation. Afterwards we apply the results to the solidification of silicon using the parameter values discussed in Sec. \ref{ssec:parameterestimation}. 

To compare the classical and non-classical heat conduction models, it will be insightful to consider the mean flux $\langle Q\rangle$, defined by
\begin{equation}
    \langle Q\rangle=\frac{1}{s}\int_0^{s}Q\,dx\, .
\end{equation}
Note, this quantity is uniform in space and, moreover, it coincides with the heat flux $Q$ when the temperature enters in the quasi-steady regime.

\subsection{MC conduction with enhanced memory}
For $\gamma,\beta\gg1$ with $\gamma = O(\beta)$ or larger, the asymptotic analysis in the Supplementary Material shows that we can split the solidification process into three time regimes. The first time regime occurs when $t=O(1)$ and captures the initial propagation of a heat wave from the cooled boundary into the seed crystal. The second time regime, $t=O(\gamma^{1/2})$, describes the transport of heat across the entirety of the seed crystal. This regime also captures the reflections of thermal waves that occur when they collide with the boundaries of the crystal. We find that the temperature waves destructively interfere, thus keeping the temperature $O(1)$ in size. However, the flux waves undergo constructive interference and thus the magnitude of the flux scales like $O(t)$.
Due to the large Stefan number, no appreciable solidification occurs during the first and second time regimes. In the third time regime, $t = O(\gamma^{1/2} \beta^{1/2})$, the temperature settles into its quasi-steady (linear) profile and solidification begins. Memory effects are relevant during the initial stages of solidification but diminish with time until Fourier's law is recovered. If $\gamma \gg 1$ but $\gamma \ll \beta$, then memory effects diminish before solidification begins.

\begin{figure}[h!]
    \centering
    \begin{subfigure}{.47\textwidth}
	\includegraphics[width=\textwidth]{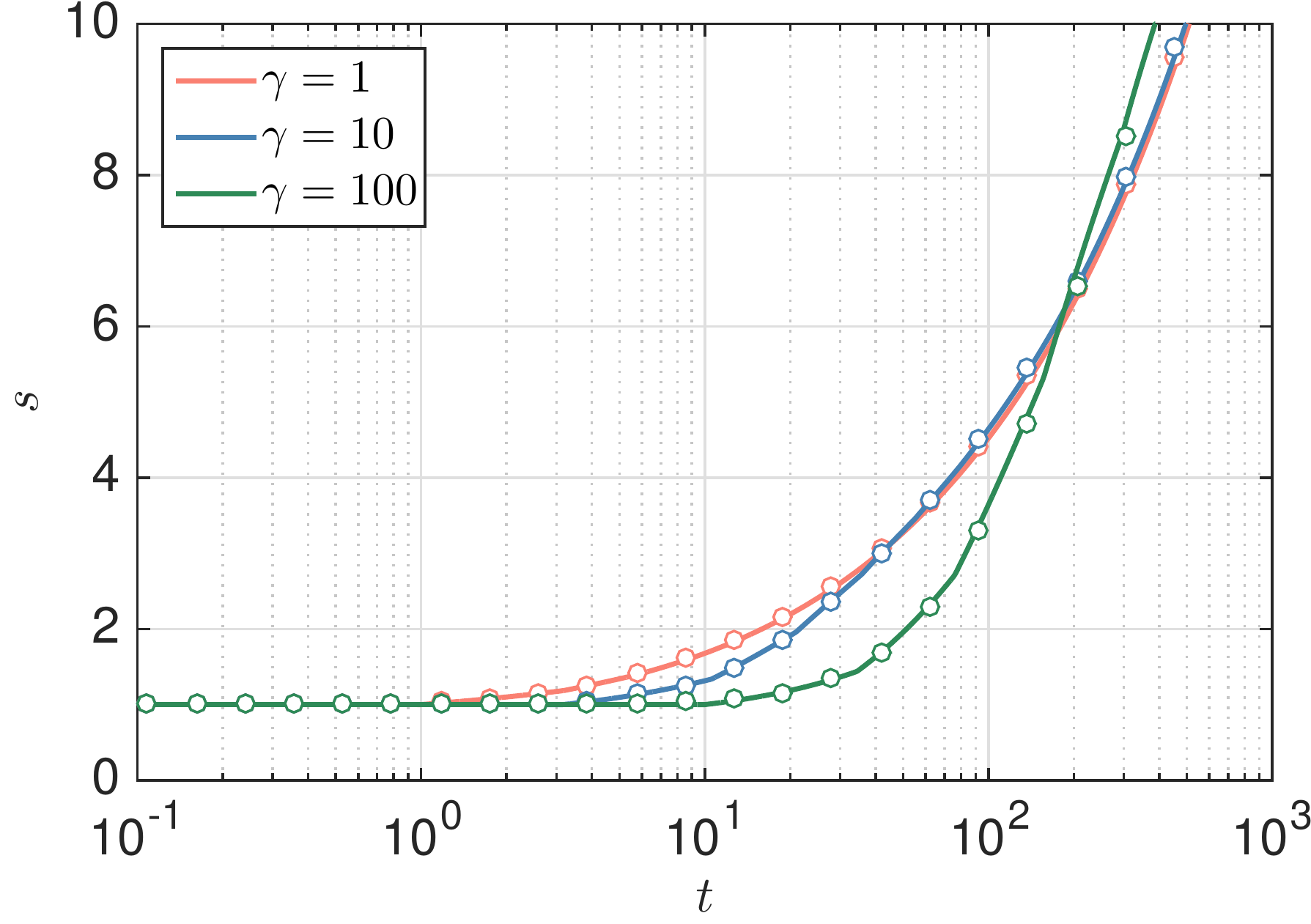}
    \caption{}\label{fig:MC_s}
	\end{subfigure}
	~
	\begin{subfigure}{.47\textwidth}
	\includegraphics[width=\textwidth]{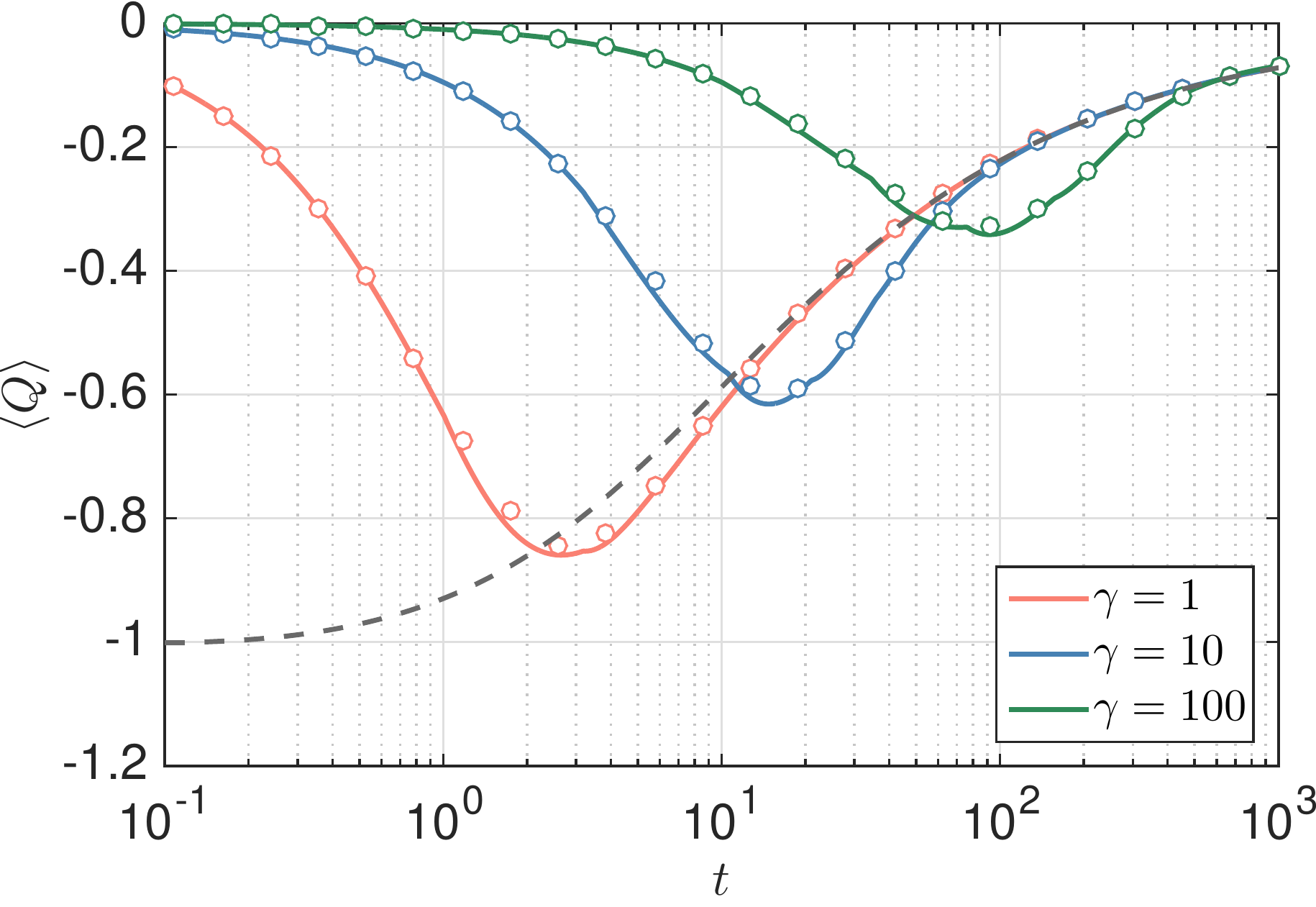}
    \caption{}\label{fig:MC_Q}
	\end{subfigure}
    \caption{Evolution of the solid-liquid interface and the mean flux according to the MCE for $\beta=10$ and different values of the Cattaneo number $\gamma$. Solid lines refer to the numerical solutions of the MC-model, whereas circles represent the solutions to the quasi-steady formulation. The dashed line in panel (b) represents the mean flux according to Fourier's law.}
    \label{fig:MC_results}
\end{figure}

Figure \ref{fig:MC_results} shows the solidification kinetics and evolution in time of the mean flux for different values of $\gamma$. Solid lines correspond to numerical simulations, whereas circles refer to the solutions of the reduced formulation. The results are shown for $\beta=10$.
Since the asymptotic analysis reveals that for $\gamma\ll\beta$ non-classical effects become negligible by the time the solid starts growing, the case $\gamma=0$ is not shown in Fig.~\ref{fig:MC_s}, as it is indistinguishable from the case $\gamma=1$. However, the mean flux for the $\gamma = 0$ case is shown in Fig.~\ref{fig:MC_Q}, where it can be clearly observed how the limit $\gamma\to0$, i.e. the Fourier case, represents a singular limit of the problem. In all cases, the agreement between the numerical simulations and the solutions of the reduced model is excellent. Simulations for larger values of $\beta$ (not shown here) also provide excellent agreement between both methods.

In Fig.~\ref{fig:MC_s} we can see how the start of the solidification process is delayed as $\gamma$ increases. This is caused by the fact that, for large values of $\gamma$, heat propagates in the form of a thermal waves with finite speed $\gamma^{-1/2}$ for a longer period and solidification begins after the wave hits the right extreme of the seed crystal when $t=O(\gamma^{1/2})$. The asymptotic analysis reveals that solidification begins when $t=O(\gamma^{1/2}\beta^{1/2})$. Hence, for $\gamma=10$ and $\gamma=100$, solidification begins when $t\approx 10$ and $t\approx32$ respectively, which is confirmed in Fig.~\ref{fig:MC_s}. The case $\gamma=100$ shows an interesting evolution of the interface. On one hand, the start of the interface motion is delayed due to the slower speed of the thermal wave through the crystal. On the other hand, the long-term rate of solidification is higher than for smaller values of $\gamma$ due to the delayed increase of the flux, which reaches its largest value when $t=O(\gamma)$, where its value is larger than the flux corresponding to smaller relaxation times. In this sense, solidification begins later as we increase $\gamma$, but the long-term solidification rate is larger until the classical solidification kinetics, where $s\sim t^{1/2}$, are recovered.

In Fig.~\ref{fig:MC_Q} we observe that, as $\gamma$ decreases, the flux converges earlier to the profile described by the classical formulation. As $\gamma$ increases, the initial MC flux decreases, in accordance with the fact that $Q=O(\gamma)$ for early times. The decrease in the mean flux from its initial value of zero is due to the repeating sequence of constructive wave reflections that occurs at the boundaries of the solid. As $\gamma$ decreases and the thermal-wave speed increases, these reflections occur more frequently, thereby leading to a more rapid decrease in the mean flux. The Fourier case can therefore be interpreted as a limit whereby these reflections occur infinitely frequently, leading to an instantaneous jump in the mean flux from its initial value of zero. The asymptotic analysis also reveals an increase of the flux for $t\gg\gamma^{1/2}$, with the order of magnitude being proportional to $(\beta/\gamma)^{1/2}$. This can be observed in Fig.~\ref{fig:MC_Q} and agrees with the asymptotic analysis. 

\subsection{GK conduction with enhanced memory}
We now assume that $\gamma,\beta\gg1$ with $\eta=O(1)$. The asymptotic analysis in the Supplementary Material reveals there are four time regimes.  When $t \ll 1$, Fourier's law is recovered with a small effective thermal conductivity (ETC) defined by $\zeta=\eta^2/\gamma$. Changes in the temperature and flux from their initial values only occur in a thin layer near $x=0$. When $t = O(1)$, heat propagates towards the liquid as a wave. The form of this wave is analogous to a viscoelastic pressure wave.  Contrary to the MC case, here the wavefront (with position given by $x_f(t)$) is not sharp but diffuse with a (dimensionless) width that scales like $w = O(\zeta^{1/2} t^{1/2})$. When $t = O(\gamma^{1/2})$ we find that the thermal and flux waves undergo a sequence of reflections at the boundaries of the solid that is analogous to the MC case, where now the width of the diffuse wavefront continuously increases. The choice of the fourth time regime depends on the relative size of $\gamma$ to $\beta$. In all cases, non-local effects become negligible, solidification begins, and the dynamics become analogous to the third time regime of the MC case.

\begin{figure}[h!]
    \centering
    \begin{subfigure}{.47\textwidth}
	\includegraphics[width=\textwidth]{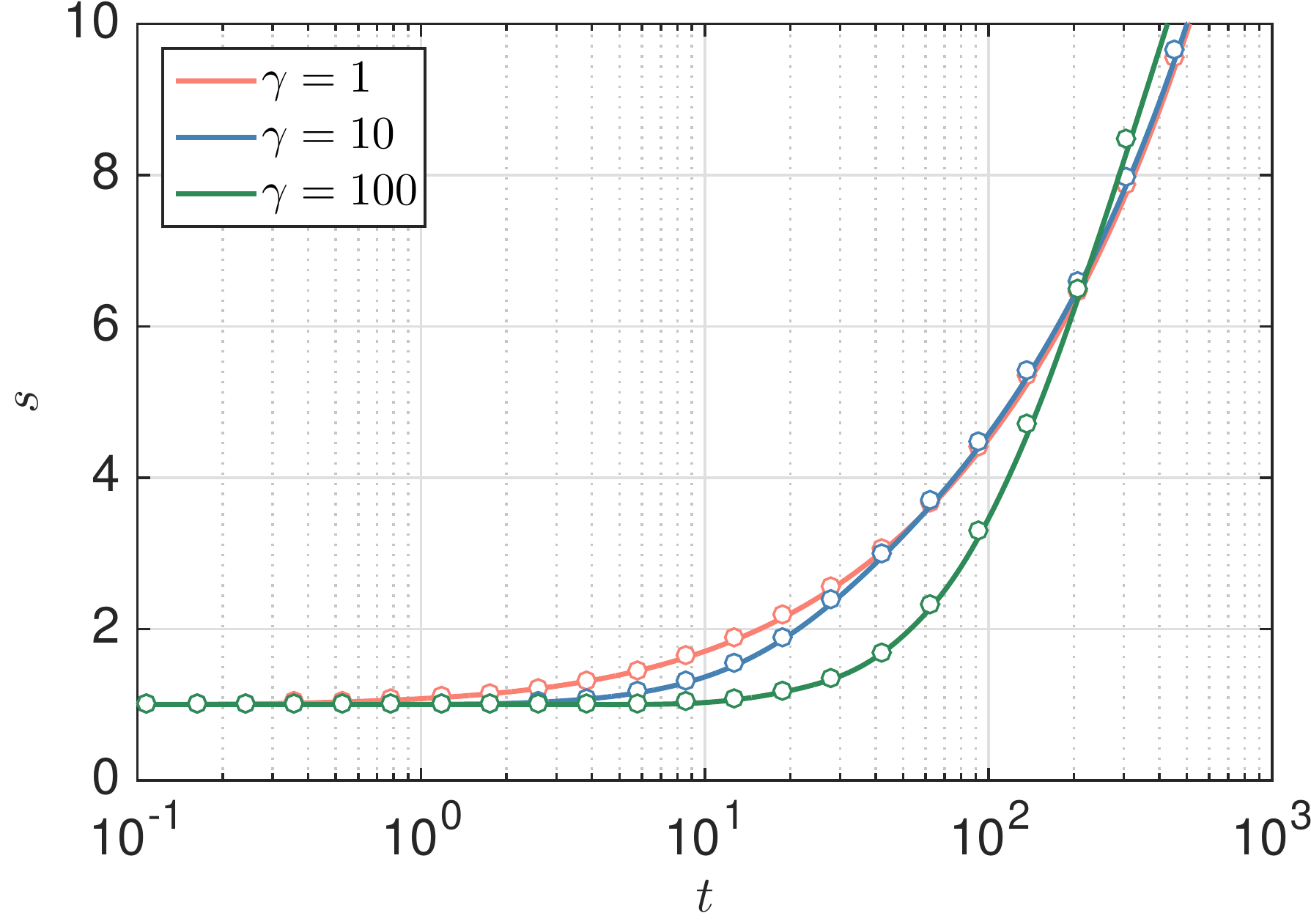}
    \caption{}\label{fig:GK_s_memory}
	\end{subfigure}
	~
	\begin{subfigure}{.47\textwidth}
	\includegraphics[width=\textwidth]{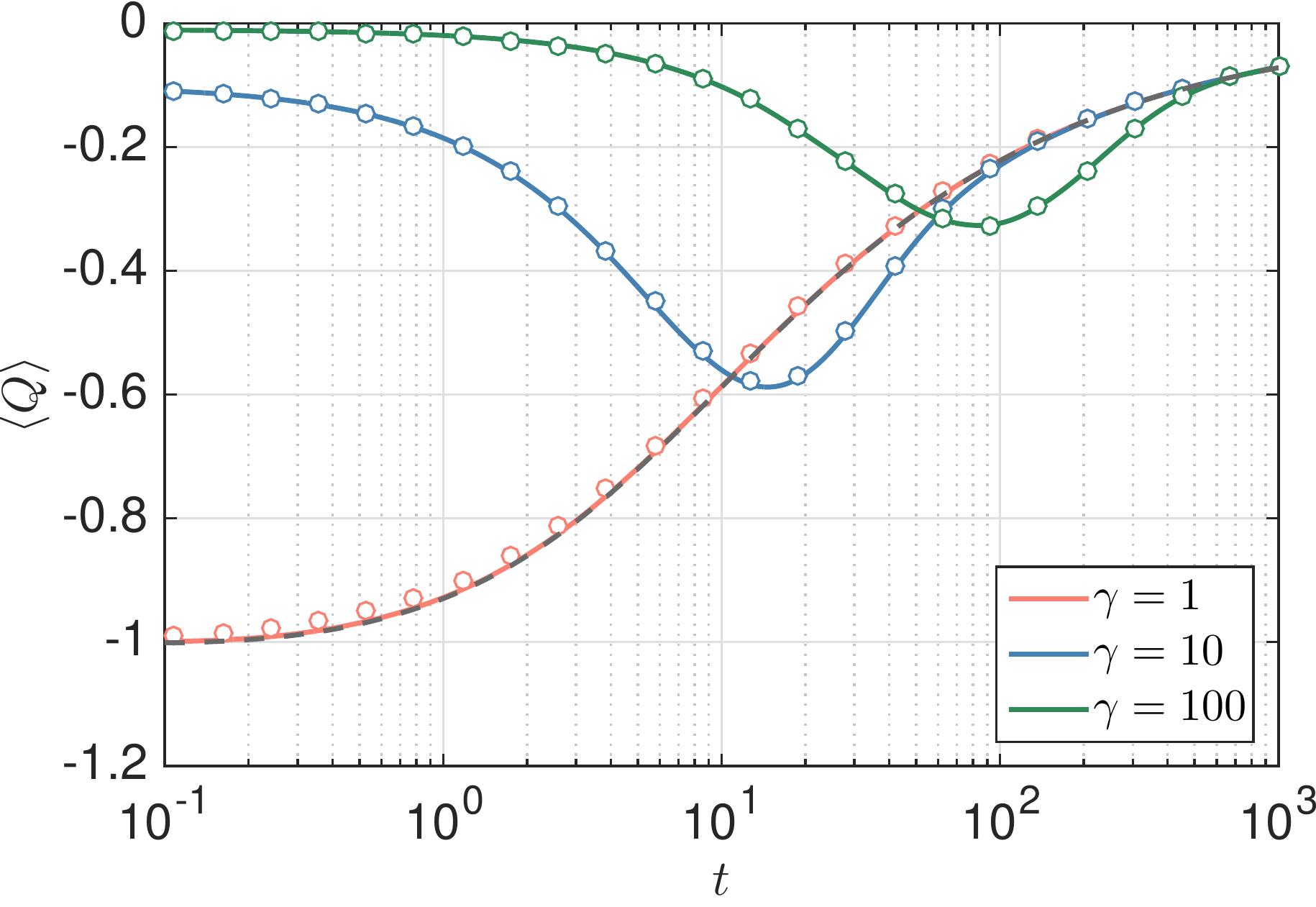}
    \caption{}\label{fig:GK_Q_memory}
	\end{subfigure}

    \caption{Evolution of the solid-liquid interface and the mean flux according to the GKE in the limit of large memory effects. The remaining dimensionless parameters are $\eta=1$ and $\beta=10$. Solid lines refer to the numerical solutions of the GK-model, whereas circles represent the solutions to the quasi-steady formulation. The dashed line in panel (b) represents the mean flux according to the classical formulation.}
    \label{fig:GK_results_memory}
\end{figure}

The evolution in time of the solid-liquid interface and the mean flux is shown in Fig.~\ref{fig:GK_results_memory} for different values of $\gamma$. The mean flux corresponding to the Fourier model is also shown (dashed, gray line). The remaining dimensionless parameters have been set to $\eta=1$ and $\beta=10$. In addition, we have plotted the case $\gamma=1$, which reproduces the classical solidification kinetics due to the Fourier resonance phenomenon, as shown by the perfect match of the corresponding fluxes in Fig.~\ref{fig:GK_Q_memory}. Similarly to the MC case, we note that the agreement between the numerical simulations and the solutions of the reduced model is remarkable.

By comparing Figs.~\ref{fig:MC_s} and \ref{fig:GK_s_memory}, it can be observed that the solidification kinetics in the GK case are very similar to those in the MC case. This was already anticipated by the asymptotic analysis, which predicts that non-local effects become negligible by the time solidification begins. In fact, the key differences between the MC and GK formulations with large relaxation times do not appear in the solidification process, since it is essentially captured by the same equations (see Secs. \ref{sec:maxwellcattaneo_third} and \ref{sec:gk:memory_fourth}), but in the initial conduction through the crystal, as can be observed by comparing Figs.~\ref{fig:MC_Q} and \ref{fig:GK_Q_memory}. However, notice that these initial differences between the MC and GK formulations disappear for $t=O(\gamma^{1/2}\beta^{1/2})$ and hence both fluxes coincide by the time solidification begins. The different behaviour for smaller times is caused by the different underlying physics in each model. The MCE predicts a wave-like heat propagation through the crystal, whereas the GKE initially recovers Fourier's law with an effective thermal conductivity $\zeta$. In the case of Fourier resonance, the ETC becomes $\zeta = 1$ and it coincides with the bulk thermal conductivity.

\begin{figure}
    \centering
    \begin{subfigure}{\textwidth}
	\includegraphics[width=.47\textwidth]{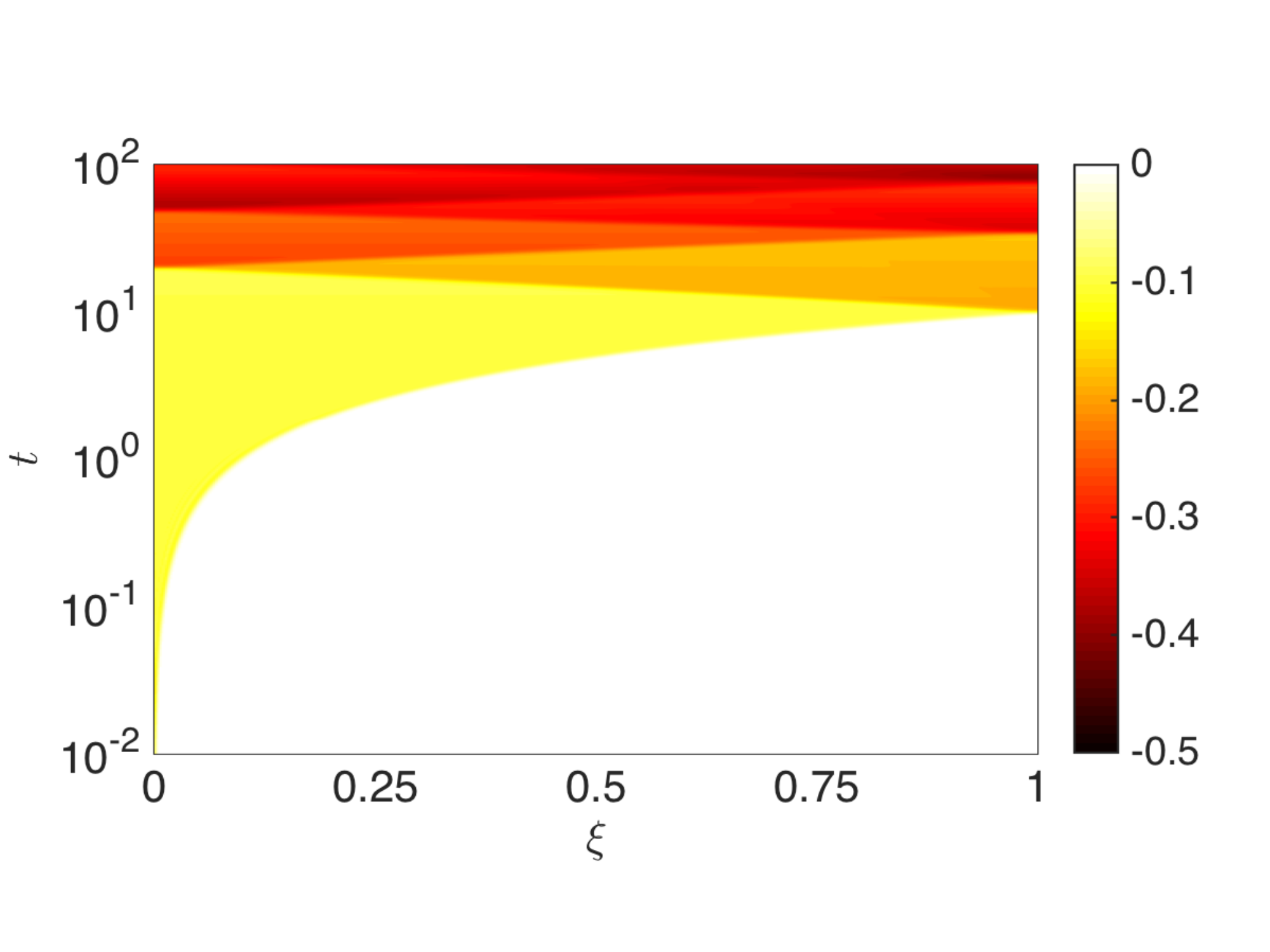}
	~
	\includegraphics[width=.47\textwidth]{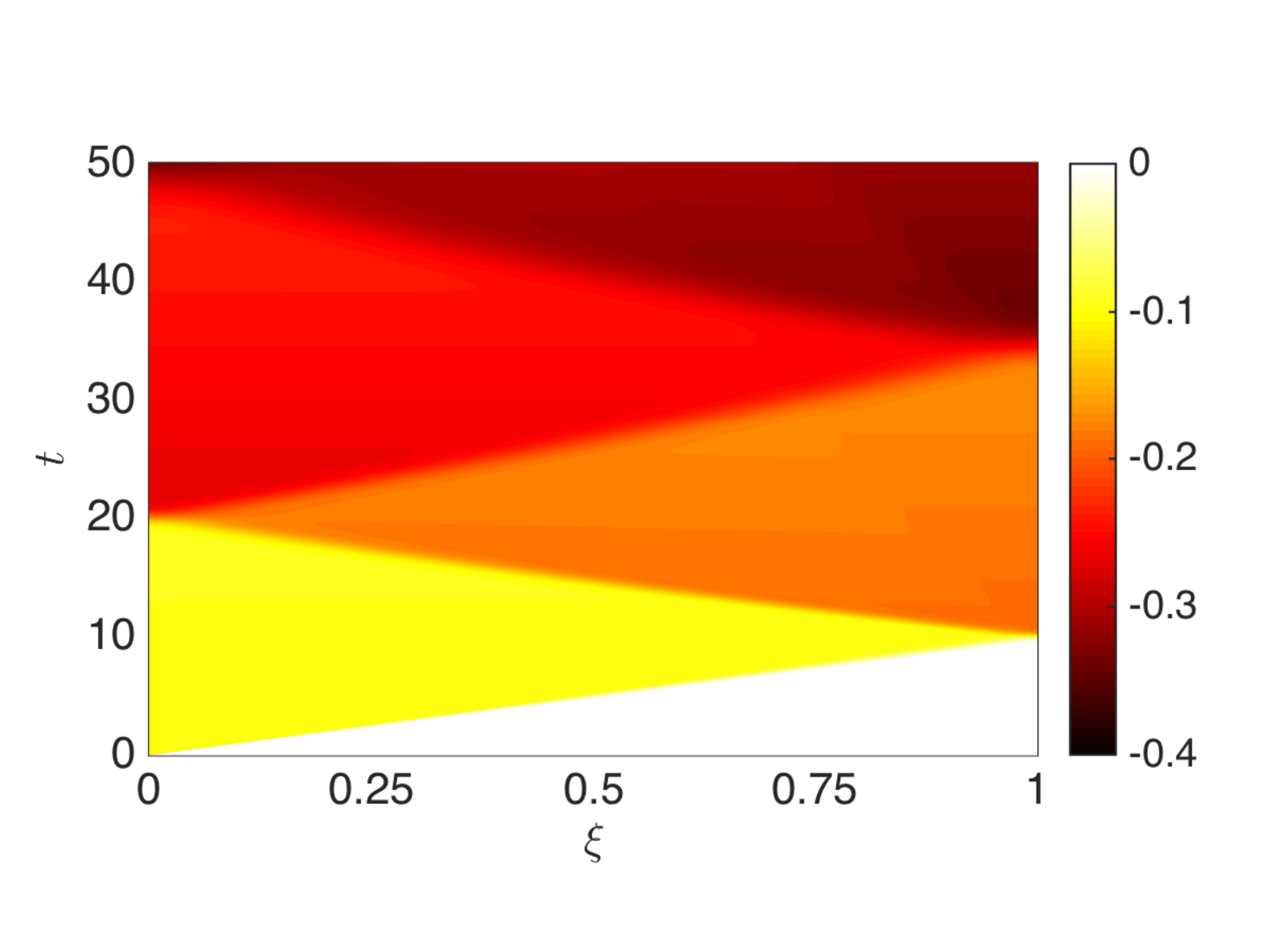}
    \caption{Maxwell-Cattaneo conduction}\label{fig:comp_fluxes_MC}
	\end{subfigure}
	
	\begin{subfigure}{\textwidth}
	\includegraphics[width=.47\textwidth]{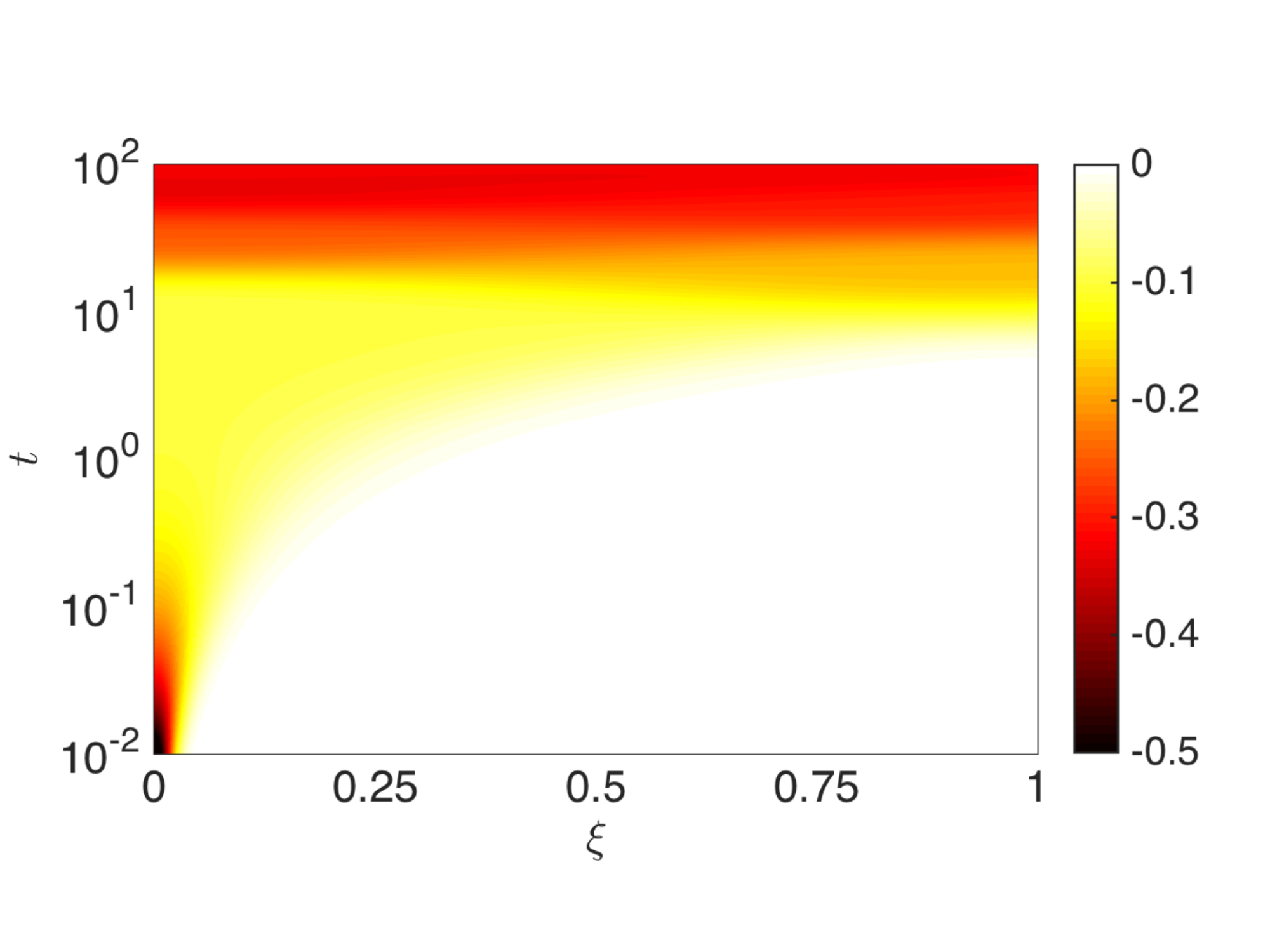}
	~	
	\includegraphics[width=.47\textwidth]{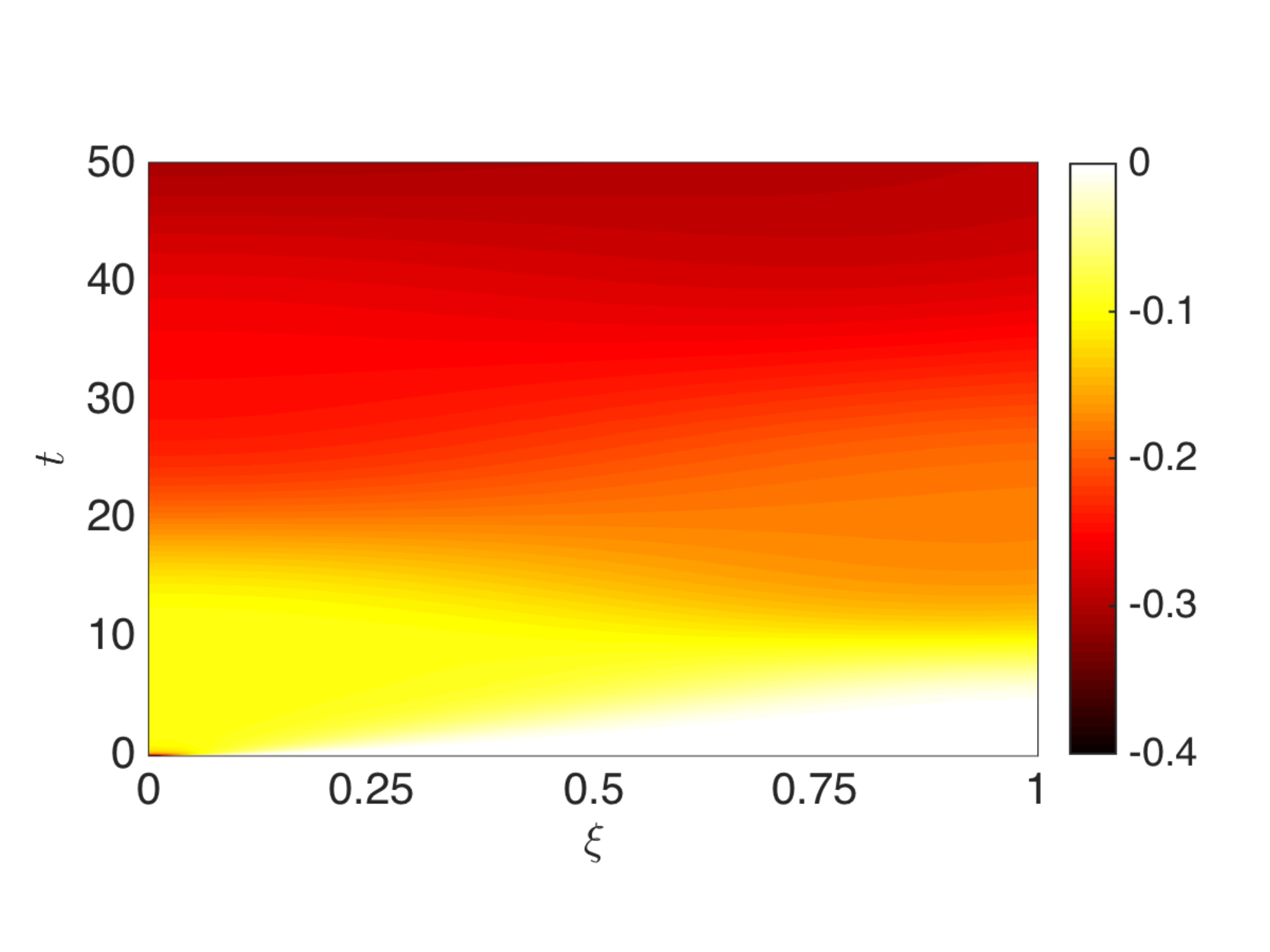}
    \caption{Guyer-Krumhansl conduction}\label{fig:comp_fluxes_GK}
	\end{subfigure}
    \caption{Heat maps showing the evolution in time of the spatial distribution of the flux in terms of the transformed variable $\xi=x/s$, according to the MCE and the GKE. For the latter we have set $\eta=1$. In both cases $\beta=10$ and $\gamma=100$. The panels in the right column focus on the time regime that captures the wave-like propagation of heat.}
    \label{fig:comp_fluxes}
\end{figure}

As revealed by the asymptotic analysis, the flux in this specific limit of the GK model is very similar to the MC model and differences appear only for small times. The evolution of the flux profiles for the MC and GK cases is shown in Fig.~\ref{fig:comp_fluxes}. To clearly illustrate the wave-like propagation of heat in the MC and GK models, we increase $\gamma$ to 100 in both cases, using $\eta = 1$ for the GK case, and fix $\beta=10$ as before. We can observe that differences in both formulations indeed appear only for small times and in the way the moving wave front $x_f(t)$ evolves in time. The right panels in Fig.~\ref{fig:comp_fluxes} show the time regimes when heat propagates in the form of thermal waves. A sharp wavefront moving at a speed $\gamma^{1/2}$ can be clearly observed in the MC case, in contrast to the GK case, which has a diffuse interface.
  
\subsection{GK conduction with enhanced non-local effects}
In this case we assume $\eta,\beta\gg1$ and $\gamma=O(1)$. From the asymptotic analysis we know that there are three time regimes to consider. In the first regime, $t=O(\eta^{-2})$,  Fourier's law is recovered with a large value of the ETC, $\zeta \gg 1$, and no solidification occurs. Due to the high value of the ETC, this regime describes the propagation of a large heat flux into the seed crystal. The second time regime depends on the relative size of $\eta^2$ to $\beta$, but in any case the system enters a quasi-steady regime. In the distinguished limit $\beta/\eta^{2}=O(1)$, the second time regime is given by $t = O(1)$, and the flux is driven by memory and non-local effects, i.e. the history and rate of change of the temperature gradient, rather than the temperature gradient itself. As time increases, the strength of these non-classical transport mechanisms diminishes and solidification decelerates, causing the position of the solid-liquid interface to reach a temporary steady state. However, as the third time regime, $t = O(\beta)$, is entered, the contributions to the flux from the temperature gradient become large enough to balance and then dominate the non-classical mechanisms, re-initiating the process of solidification. For large times, the classical solidification kinetics, $s \sim t^{1/2}$, are recovered.  Similar time regimes were observed in Hennessy et al.~\cite{Hennessy2018}, although in their case the magnitude of the flux for small times is not as large as here due to their use of a Newton condition at $x = 0$ rather than a fixed-temperature condition.

\begin{figure}[h!]
    \centering
    \begin{subfigure}{.47\textwidth}
	\includegraphics[width=\textwidth]{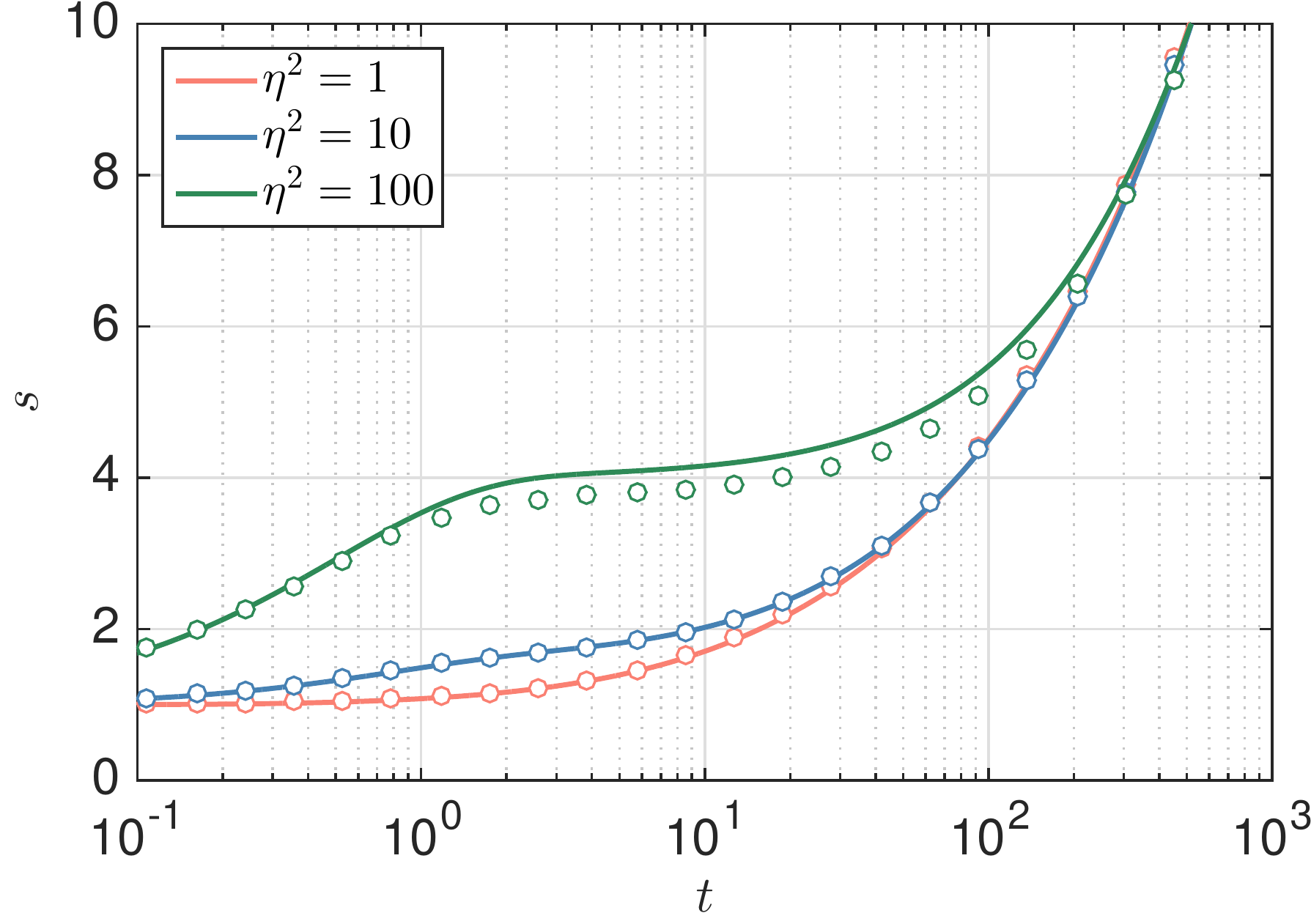}
    \caption{}\label{fig:GK_s_nonlocal}
	\end{subfigure}
	~
	\begin{subfigure}{.47\textwidth}
	\includegraphics[width=\textwidth]{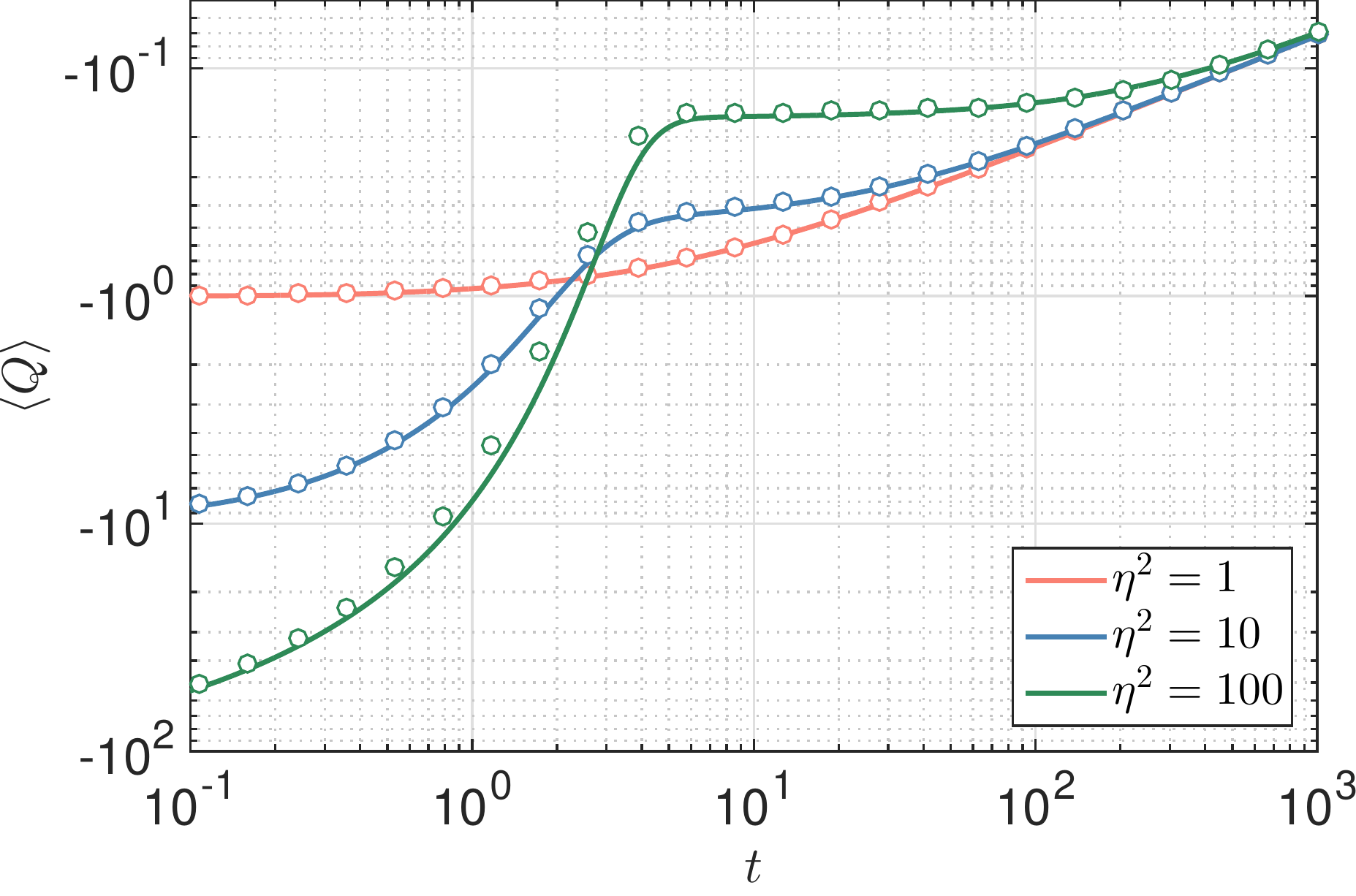}
    \caption{}\label{fig:GK_Q_nonlocal}
	\end{subfigure}
    \caption{Evolution of the solid-liquid interface and the mean flux according to the GKE in the limit of large non-local effects. The remaining dimensionless parameters are $\gamma=1$ and $\beta=10$. Lines refer to the numerical simulations, whereas circles represent the solutions to the quasi-steady formulation.}
    \label{fig:GK_results_nonlocal}
\end{figure}

The evolution of the interface and the mean flux is shown in Fig.~\ref{fig:GK_results_nonlocal}, where lines and circles again represent solutions computed numerically and via the reduced formulation. Due to the large variation of the magnitude of $\langle Q\rangle$, the results in Fig.~\ref{fig:GK_Q_nonlocal} are shown using logarithmic scales on both axes. As in the case of enhanced memory effects, we also show the results for $\eta=1$, which reduces to the classical model due to the Fourier resonance phenomenon. In general, there is good agreement between the solution to the reduced formulation and numerical simulations, with some small discrepancies in Fig.~\ref{fig:GK_s_nonlocal} that may be attributed to the terms being neglected in the asymptotic analysis. Larger discrepancies may be observed in Fig.~\ref{fig:GK_Q_nonlocal} for $t\ll1$, where the reduced formulation predicts smaller values (in magnitude) of the heat flux than the numerical simulations. This error is due to the reduced model not capturing these small-time regimes.

As predicted by the asymptotic analysis, the initial solidification rate increases as the non-local parameter $\eta$ is increased. This is attributed to the increase in the ETC, which enhances the transport of thermal energy to the solidification front. The rapid initial growth of the solid is solely driven by non-Fourier conduction mechanisms; the contribution to the flux from the instantaneous temperature gradient is negligible for small times. As time increases and the solid grows in size, memory and non-local effects weaken and the solidification rate process undergoes a substantial decrease, as seen in Fig.~\ref{fig:GK_results_nonlocal} when $\eta^2 = 100$. The mean flux relaxes to roughly the Fourier contribution, $\langle Q \rangle \simeq -1/s$; however, due to the rapid initial growth of the solid ($s \gg 1$), this contribution is strongly diminished compared to the case of pure Fourier conduction. As a result, when solidification is re-initiated on $O(\beta)$ time scales, it proceeds at a greatly reduced rate.

In summary, the dynamics that occur in the case of GK conduction with strong non-local effects are the opposite of those that occur in the limit of strong memory effects or for MC conduction. In the former case $\eta \gg 1$ with $\gamma = O(1)$, the mean thermal flux is initially much greater and then much smaller than the Fourier prediction, which leads to solidification kinetics that are much faster and then much slower than the Fourier case. However, if $\gamma \gg 1$ with $\eta = O(1)$ (or $\eta = 0$), then there is a prolonged period where the mean thermal flux is initially much higher and then slightly less than the Fourier prediction, resulting in relatively slower and then faster solidification kinetics.

  \subsection{Application to the solidification of Si}

\begin{figure}[h!]
    \centering
    \begin{subfigure}{.47\textwidth}
	\includegraphics[width=\textwidth]{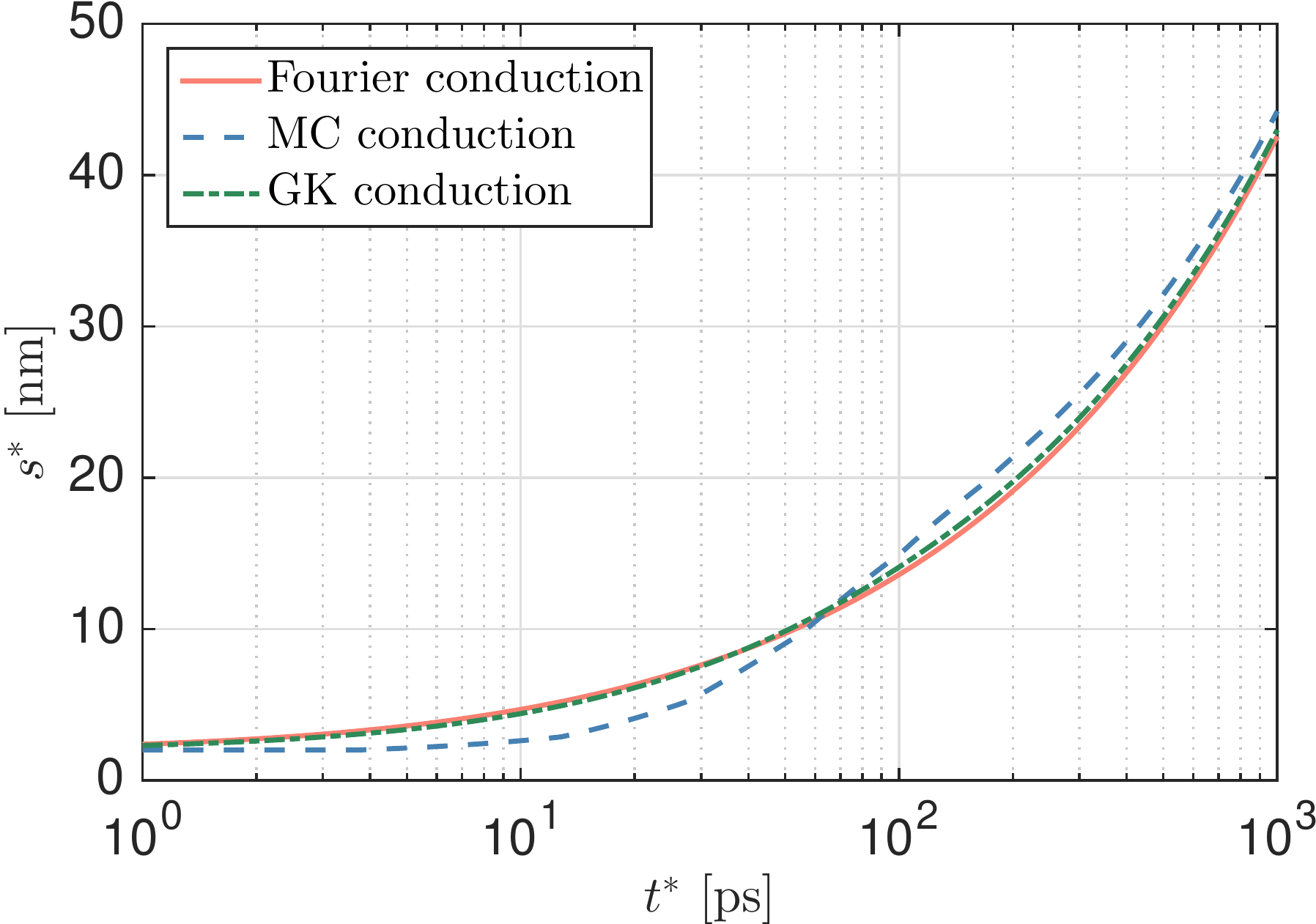}
    \caption{$s_c^*=2$ nm}\label{fig:results_Si_a}
	\end{subfigure}
	~
	\begin{subfigure}{.47\textwidth}
	\includegraphics[width=\textwidth]{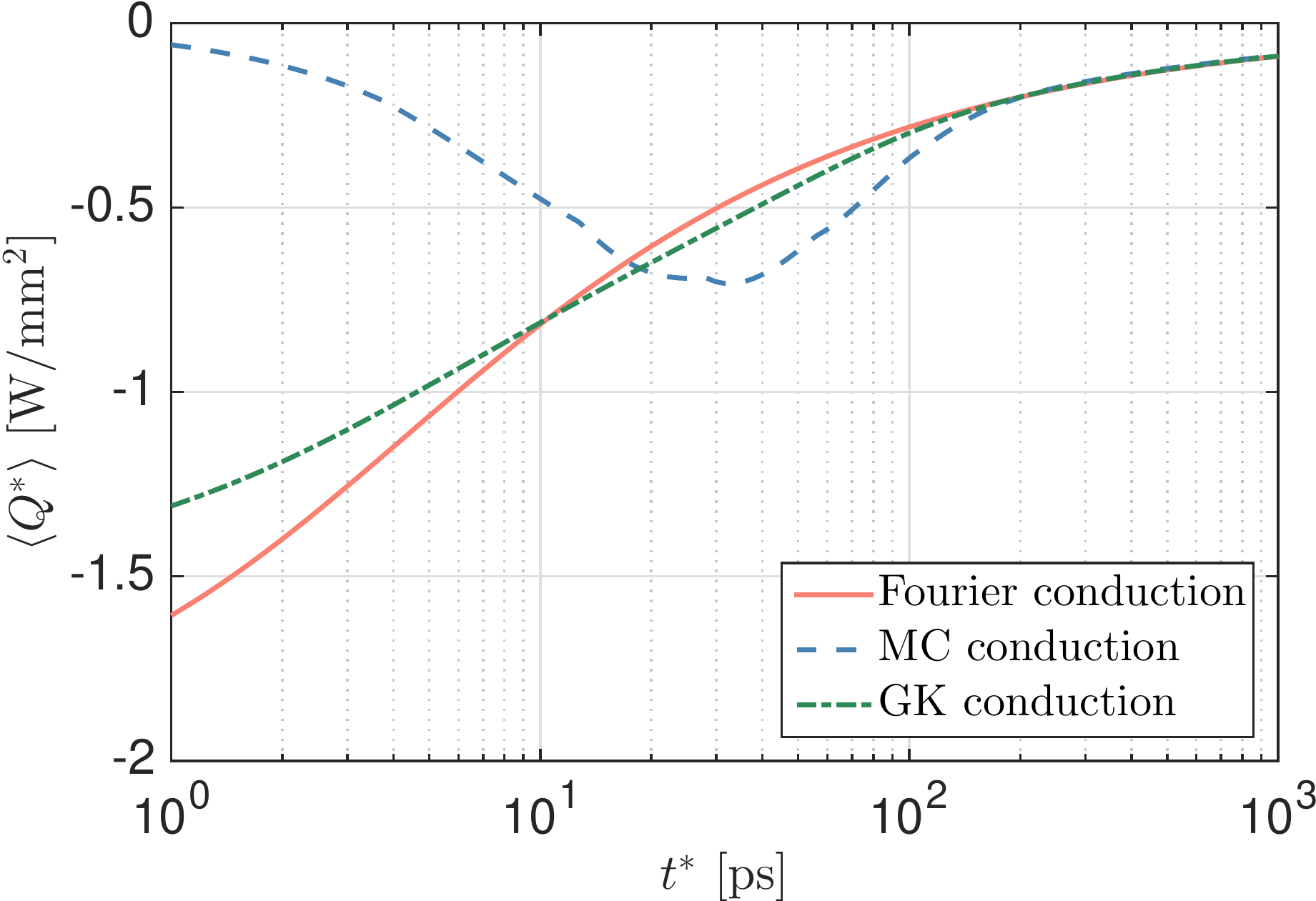}
    \caption{$s_c^*=2$ nm}\label{fig:results_Si_b}
	\end{subfigure}
	
	\begin{subfigure}{.47\textwidth}
	\includegraphics[width=\textwidth]{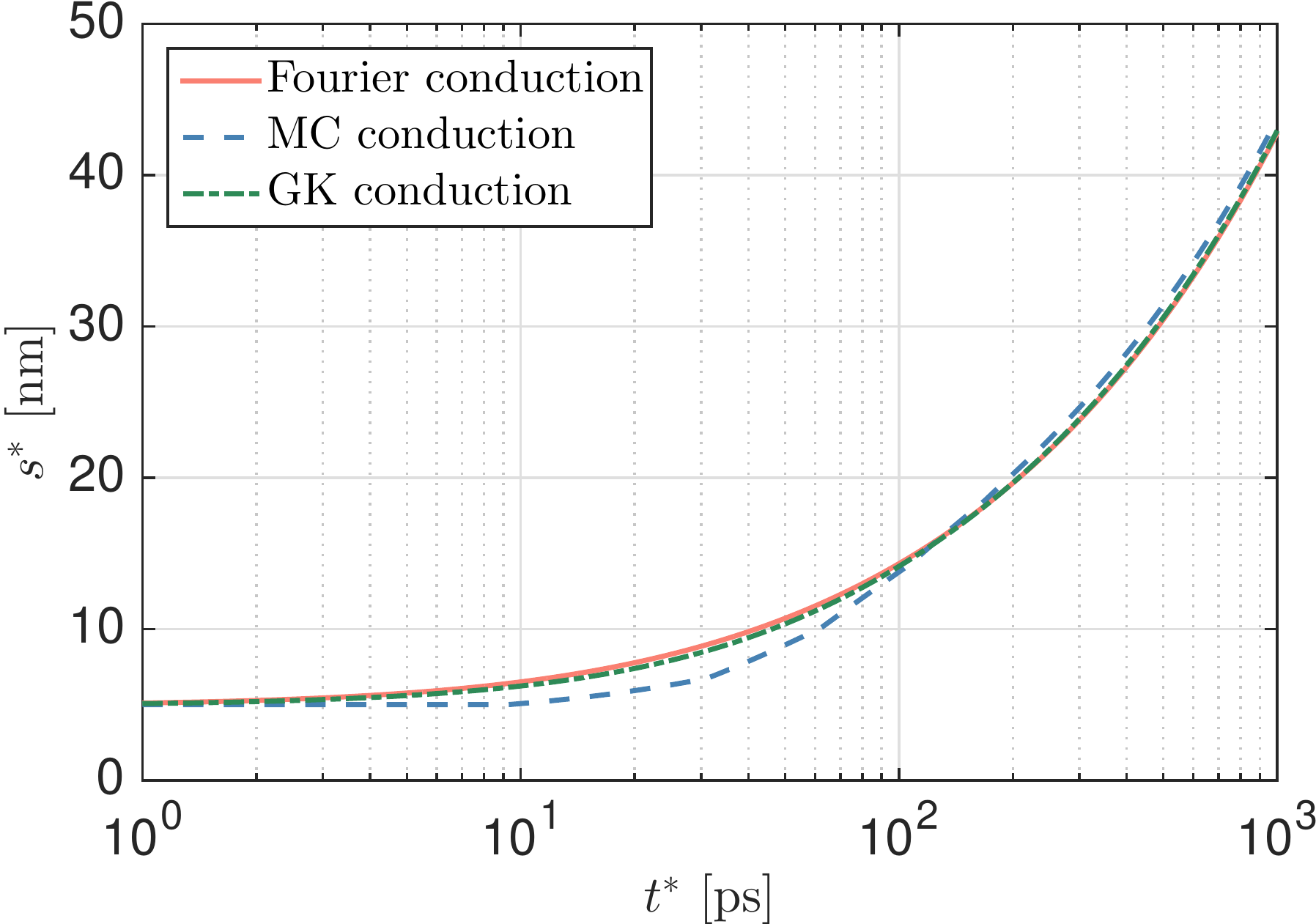}
    \caption{$s_c^*=5$ nm}\label{fig:results_Si_c}
	\end{subfigure}
	~
	\begin{subfigure}{.47\textwidth}
	\includegraphics[width=\textwidth]{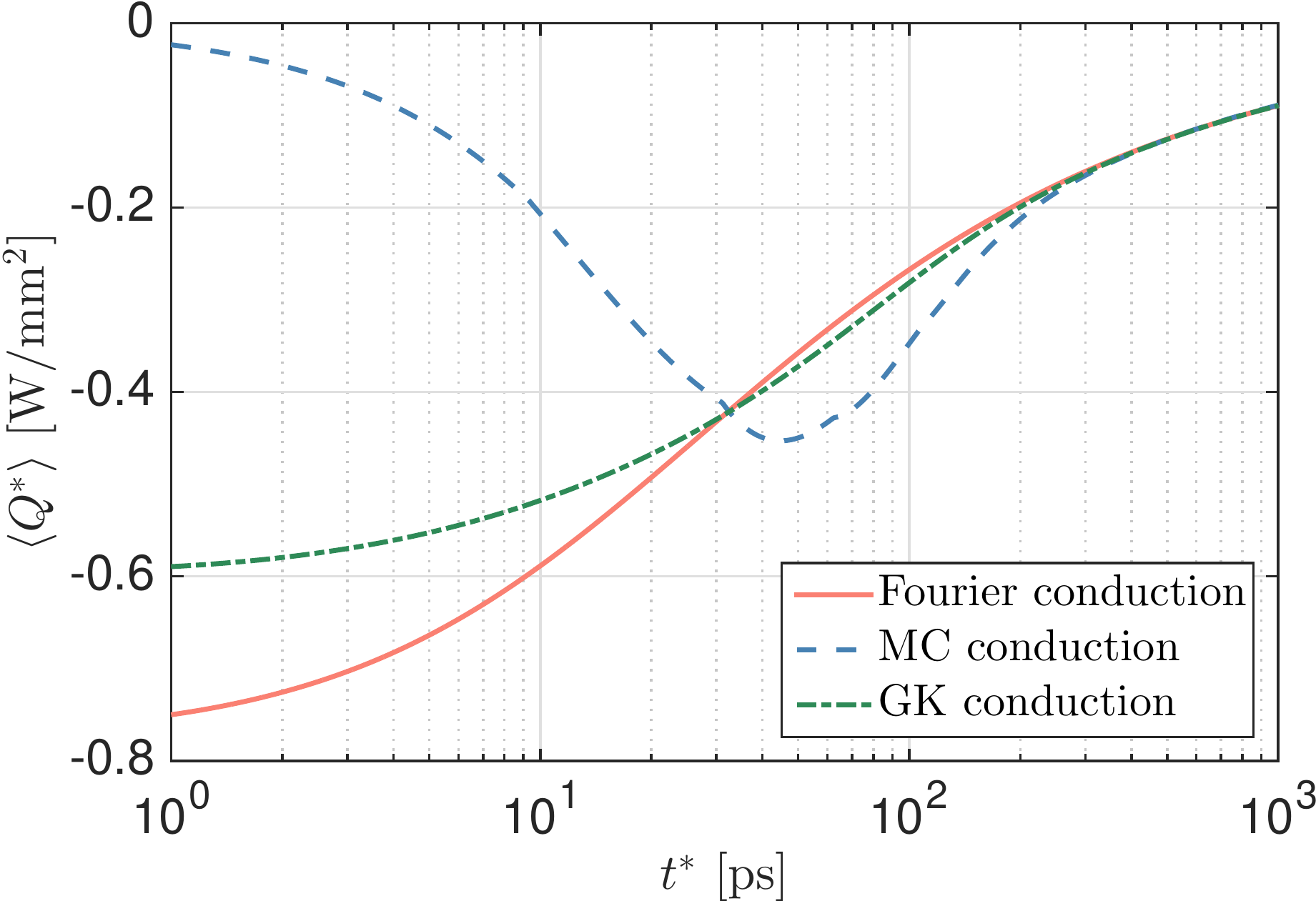}
    \caption{$s_c^*=5$ nm}\label{fig:results_Si_d}
	\end{subfigure}
	
	\begin{subfigure}{.47\textwidth}
	\includegraphics[width=\textwidth]{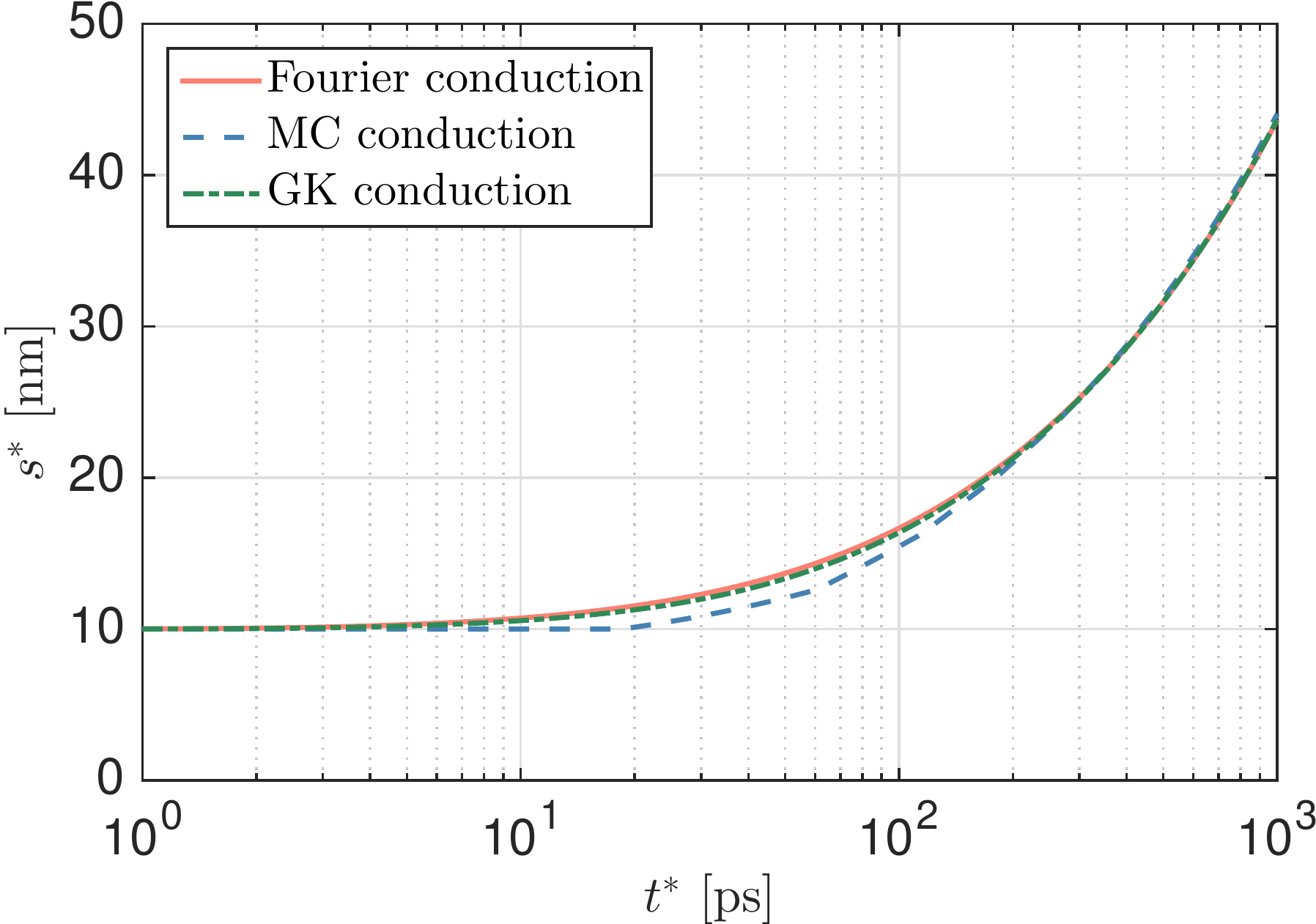}
     \caption{$s_c^*=10$ nm}\label{fig:results_Si_e}
	\end{subfigure}
	~
	\begin{subfigure}{.47\textwidth}
	\includegraphics[width=\textwidth]{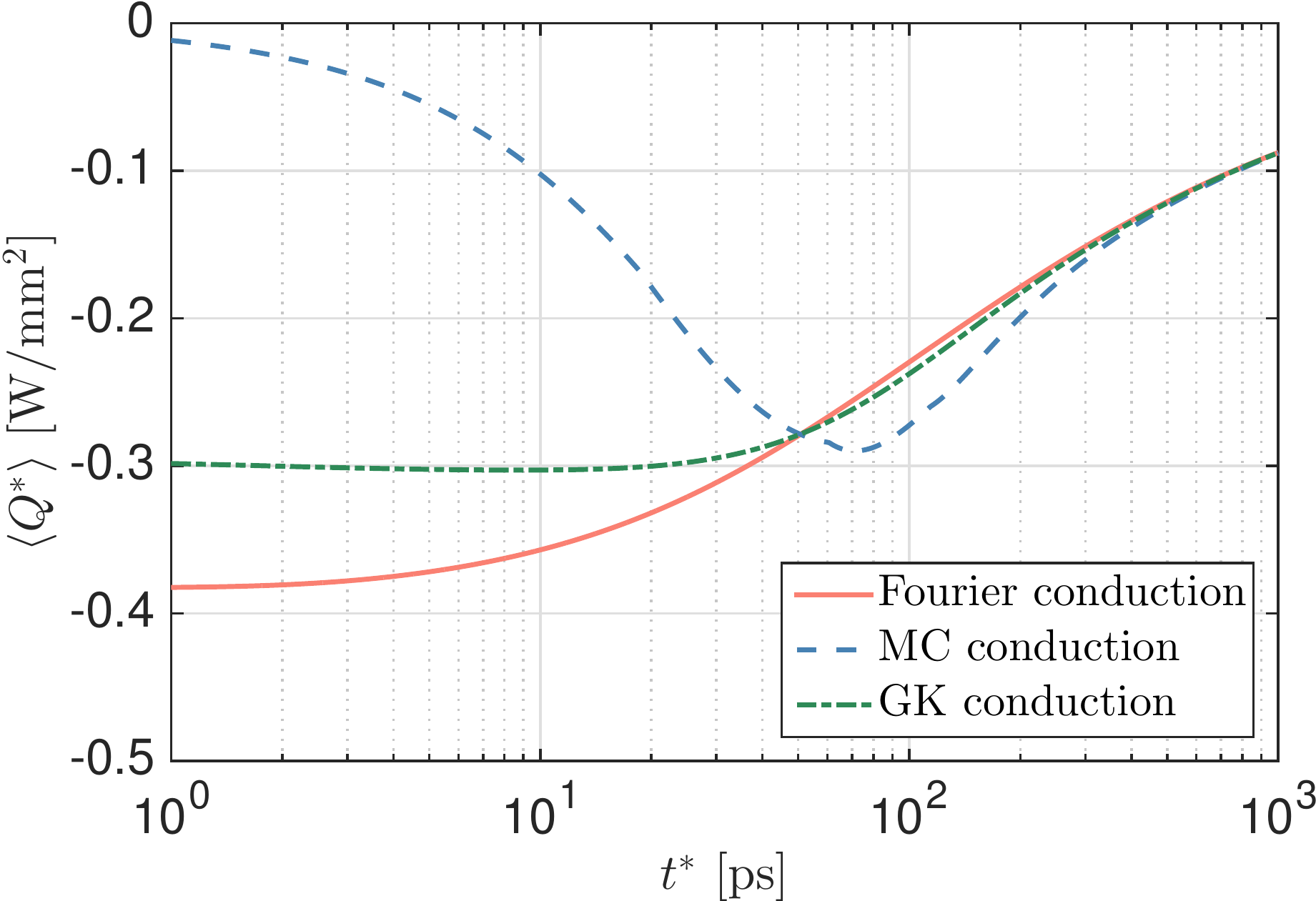}
    \caption{$s_c^*=10$ nm.}\label{fig:results_Si_f}
	\end{subfigure}
    \caption{Evolution of the solid-liquid interface and the mean flux according to Fourier's law (red, solid lines), the MCE (blue, dashed lines) and the GKE (green, dashed-dotted lines) for $\beta=10$ and different sizes of the seed crystal. The values of $\gamma$ and $\eta$ are given in Table~\ref{table:ndparams_silicon}.}
    \label{fig:results_Si}
\end{figure}

  Silicon is a material of high interest for nanoscale applications and an obvious choice for the validation of theoretical models. In Fig.~\ref{fig:results_Si} we plot the evolution of the solid-liquid interface and the mean flux according to the classical, MC and GK formulations for different sizes of the seed crystal. The corresponding non-classical parameters are given in Table~\ref{table:ndparams_silicon}. The results have been calculated numerically for $\beta=10$, which corresponds to an initial temperature jump $\Delta T^*=173.2$ K. The case $\beta=100$ ($\Delta T^*=17.3$ K) leads to all curves in Figs.~\ref{fig:results_Si_a}, \ref{fig:results_Si_c} and \ref{fig:results_Si_e} collapsing and thus no differences between classical and non-classical solidification kinetics can be observed. To give the results in dimensional form, we have written $\langle Q^*\rangle=Q^*_0\langle Q\rangle$ and $t^*=\tau^*_D t$, with $Q^*_0$ and $\tau^*_D$ given in Table~\ref{table:ndparams_silicon} for each of the seed crystal sizes.

\begin{table}[t]
    \centering
    \caption{Values of the time and flux scales and the dimensionless parameters $\gamma$ and $\eta^2$ depending on the size of the seed crystal. The Stefan number has been fixed to 10.}
    \label{table:ndparams_silicon}
    \begin{tabular}{ccccc}
        \hline
        $s_c^*$ [nm] & $\tau^*_D$ [ps] & $Q^*_0$ [W/mm\units{2}] & $\gamma$  & $\eta$ \\ \hline\hline
         2   & 0.43  & 1.9 & 75 & 7.7\\
         5   & 2.7  & 0.77  & 12 & 3.0\\
         10  & 11 & 0.38  & 3.0  & 1.5\\
         \hline
    \end{tabular}
\end{table}

Figures \ref{fig:results_Si_a}, \ref{fig:results_Si_c} and \ref{fig:results_Si_e}, where we plot the evolution of the solid-liquid interface, show that the presence of non-classical effects does not cause any major differences in the solidification kinetics. In fact, the Fourier and GK models lead to virtually identical behaviours in the evolution of the solid-liquid interface for the three seed crystal sizes considered. In the case of the MC model, there is a delay in the start of the solidification, which from a physical point of view can be understood from the fact that heat is initially propagated in the form of thermal waves with finite speed, contrary to the Fourier and GK models. Furthermore, the asymptotic analysis of the MC model reveals that in the cases $s_c^*=2$ nm and $s_c^*=5$ nm, where $\gamma$ is relatively large, solidification starts for $t^*\approx11.78$ ps and $t^*\approx29.36$ ps respectively, which is in good agreement with Fig.~\ref{fig:results_Si_a}. 

Although differences in the plots corresponding to $s^*$ are small, Figs.~\ref{fig:results_Si_b}, \ref{fig:results_Si_d} and \ref{fig:results_Si_f}, which show the evolution in time of the mean flux, suggest that the choice of the underlying model does have a crucial impact on how heat will propagate through the crystal for small times ($<100$~ns). It can be observed how the qualitative behaviour of the Fourier and GK fluxes is similar whereas they have large differences with respect to the MC flux. These differences are a direct consequence of the different behaviours for small times, since in the latter heat is initially propagated in the form of heat waves whereas for the GK and classical models the initial heat propagation corresponds to a diffusive process. Therefore,  the results shown in Fig.~\ref{fig:results_Si} suggest that non-classical mechanisms can be detected by measuring the total heat flux through the solid.

The similarities between the solidification kinetics of the Fourier and GK models could exist for two reasons. Firstly, the asymptotic analysis shows non-classical effects become less important for the solidification process as the Stefan number increases. Secondly, the ratio $\zeta=\eta^2/\gamma=3\ell^{*2}/(\alpha^*\tau^*_R)$ does not depend on the initial size of the seed crystal. Using the values provided in Table~\ref{tab:silicon}, for silicon we find $\zeta\approx0.77$, which implies that the similarities may be caused by the system being close to the case of Fourier resonance ($\zeta = 1$).

\begin{figure}[h!]
    \centering
	\begin{subfigure}{.47\textwidth}
	\includegraphics[width=\textwidth]{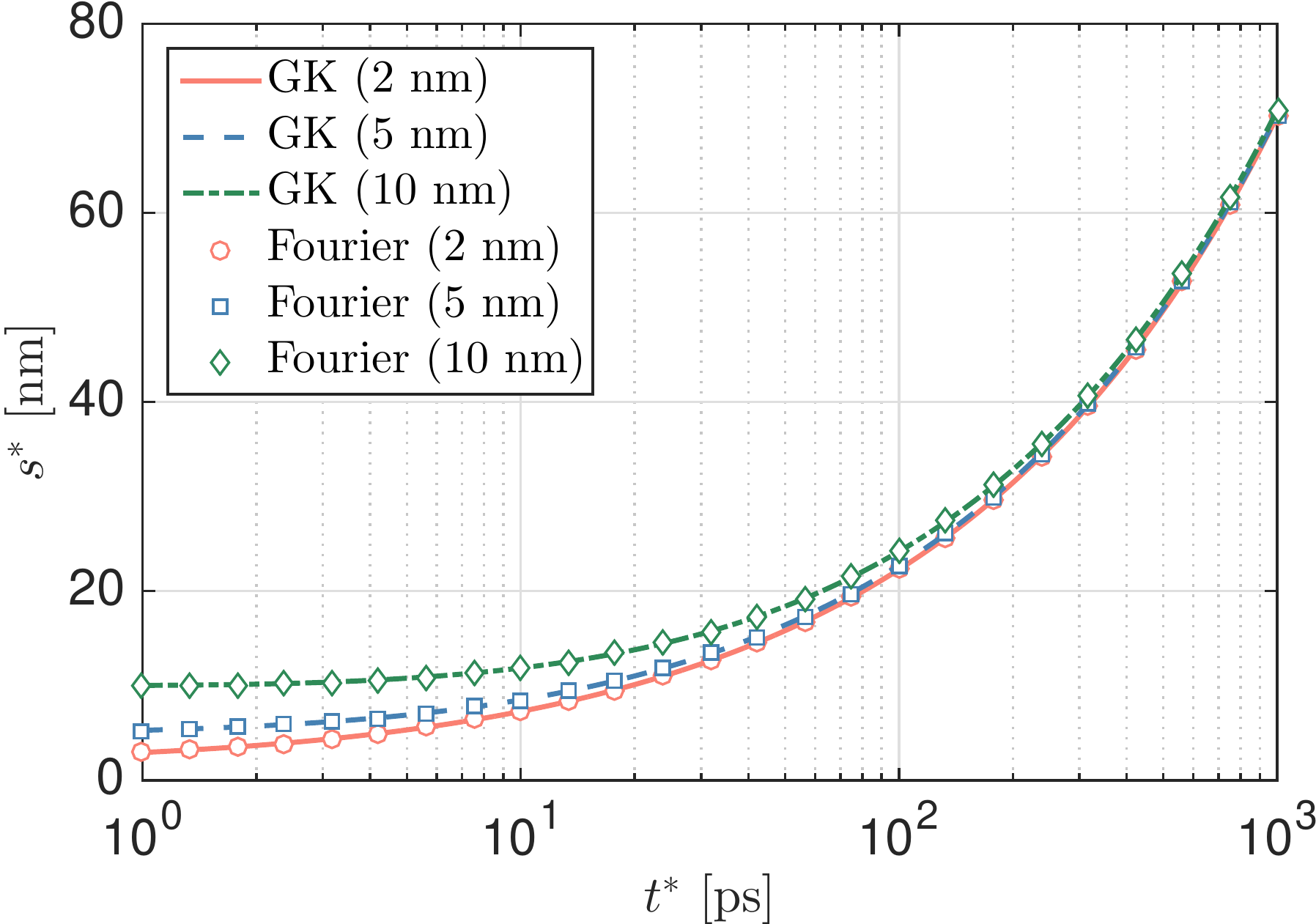}
    \caption{}\label{fig:comp_GKvsFourier}
	\end{subfigure}
	~
	\begin{subfigure}{.47\textwidth}
	\includegraphics[width=\textwidth]{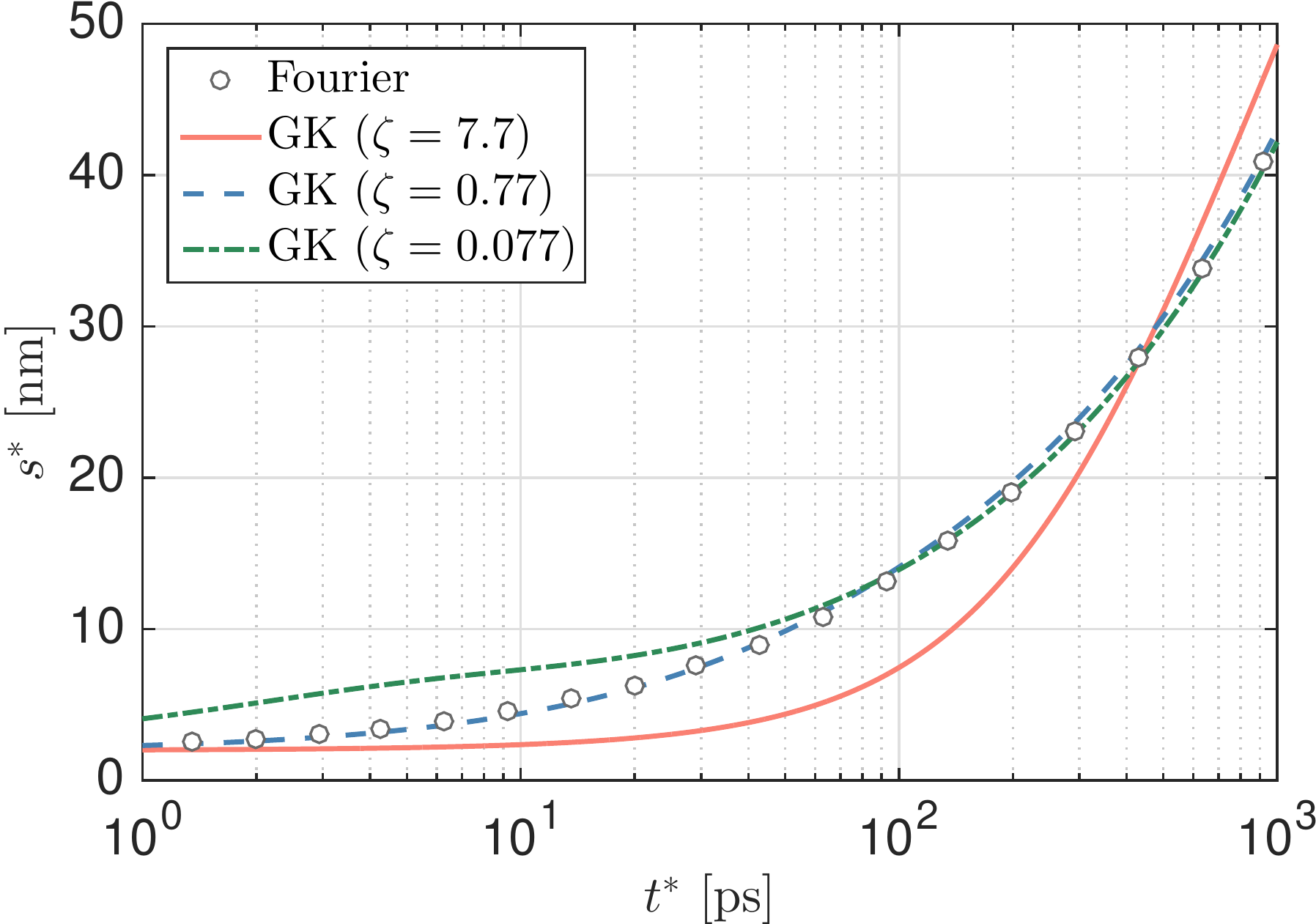}
    \caption{}\label{fig:Fourier_resonance}
	\end{subfigure}	
    \caption{(a) Evolution of the solid-liquid interface according to GK (lines) and Fourier (symbols) models for different initial seed crystal sizes and a temperature jump $\Delta T^*=500$ K ($\beta\approx3.46$). (b) Evolution of the solid-liquid interface according to the GK (lines) and Fourier (circles) models for an initial seed crystal of size $s_c^*=2$ nm and a temperature jump $\Delta T^*=173.16$ K ($\beta=10$). The solutions plotted in solid (red), dashed (blue) and dashed-dotted (green) lines correspond to the  GK model using $0.1\cdot\tau^*_R$, $\tau^*_R$ and $10\cdot\tau^*_R$, where $\tau^*_R$ is the extrapolated value given in Table~\ref{tab:silicon}.}
    \label{fig:results_Si_2}
\end{figure}

In Fig.~\ref{fig:comp_GKvsFourier} we show the solidification kinetics for three different seed crystal sizes and for a larger temperature jump of $\Delta T^*=500$ K, which reduces the Stefan number to $\beta\approx3.46$. The results show that the similarities between the Fourier and GK models persist, which indicates that the largeness of the Stefan number was not causing the similarities in the solidification kinetics observed in Figs.~\ref{fig:results_Si_a}, \ref{fig:results_Si_c} and \ref{fig:results_Si_e}.

Thus, we now explore whether these similarities are caused by Fourier resonsance phenomenoma by varying the size of the ETC $\zeta$. In Fig.~\ref{fig:Fourier_resonance} we show how the solidification kinetics vary when changing the order of magnitude of $\tau_R^*$ while keeping the value of $\ell^*$ fixed, which leads to variations in $\zeta$ across two orders of magnitude. In the cases where $\zeta\ll1$ or $\zeta\gg1$, the non-Fourier signatures predicted by the asymptotic analysis can be observed. This shows that the similarities between the solidification kinetics in the Fourier and GK models are indeed caused by the fact that $\zeta$ is close to unity. Hence, evidence of non-classical heat transport can be found in measurements of the solidification front only in the cases $\zeta\gg1$ or $\zeta\ll1$. 

From the results of this study we can conclude that non-classical conduction mechanisms do not lead to significant changes in the solidification kinetics in situations near Fourier resonance, despite the fact that the dimensionless parameters characterising these effects are large. Hence, in this case non-classical heat transport mechanisms are unlikely to be detected from measurements of the solidification front. However, the analyses predict that large deviations can occur in the thermal flux, which is a better quantity to use in the detection of non-classical transport mechanisms and which can be measured both experimentally and in molecular dynamics simulations \cite{Font2018}.

\section{Conclusion}\label{sec:conclusion}
In this paper we have studied non-classical heat transport in phase-change processes as described by the Maxwell-Cattaneo (MC) and Guyer-Krumhansl (GK) equations via the solidification kinetics of a one-dimensional liquid bath. Based on a detailed asymptotic analysis we proposed a simplified formulation which reduces the moving boundary problem to a set of two ordinary differential equations for the position of the interface and the mean flux through the solid. Comparisons of the results provided by this reduced formulation against numerical simulations have shown excellent agreement. When the non-classical terms become dominant, the study shows that we can observe large deviations from the classical solidification kinetics described by Fourier's law. However, in GK conduction near Fourier resonance, for example with silicon, the non-classical effects almost cancel each other out. This leads to similar evolution of the phase-change front between the GK and Fourier models. However, the propagation of heat through the solid is affected by the choice of the constitutive equation to a greater extent. We conclude that a possible strategy to determine the presence of non-classical effects is studying the heat conduction through the growing solid rather than the position of the interface, since the prior seems to give more information even in cases that are close to Fourier resonance. Moreover, due to the different small-time behaviours, measuring the flux could also indicate which of the considered models is more realistic, since the MC model does not have Fourier resonance. Although we focused on MC and GK conduction laws, the reduction framework developed here can be applied to alternative non-Fourier models based on continuum theory and used to study other modes of phase change such as melting as well.

\section*{Acknowledgements}
M.C.S. acknowledges that the research leading to these results received funding from ``La Caixa" Foundation and the Ferran Sunyer i Balaguer Foundation.  M.G.H and T.G.M received funding from the European Union's Horizon 2020 research and innovation programme under the Marie Sk{\l}odowska-Curie grant agreement No. 707658.  T.G.M. acknowledges the support of a Ministerio de Ciencia e Innovaci\'{o}n grant MTM2017-82317-P.  The authors have been partially funded by the CERCA programme of the Generalitat de Catalunya. The authors thank F.~X. Alvarez and L. Sendra for providing the KCM data.

\bibliographystyle{unsrtnat}
\bibliography{Refs_NFS}


\newpage

\begin{center}\huge{\textbf{Supplementary Material}}\end{center}
\beginsupplement

\normalsize

\linespread{1.3}

\section{Introduction}
In this Supplementary Material we focus on the numerical solution and asymptotic analysis of the Maxwell-Cattaneo-Stefan and the Guyer-Krumhansl-Stefan problems. The relevant, dimensionless equations are
\begin{subequations}\label{supp:eqs}
\begin{alignat}{3}
    &\gamma Q_t+Q=-T_x+\eta^2Q_{xx},\qquad && 0<x<s,\label{supp:gk}\\
    &T_t+Q_x=0,\qquad && 0<x<s,\label{supp:ce}\\
    &T=-1,\qquad &\text{at }& x=0,\label{supp:bc_0}\\
    &T=0,\qquad &\text{at }& x=s,\label{supp:bc_s}\\
    &\beta s_t=-Q,\qquad &\text{at }& x=s,\label{supp:stefan}\\
    &T=Q=0, \quad s=1,\qquad &\text{at }& t=0.\label{supp:ic}
\end{alignat}
\end{subequations}
For $\eta=0$ the GKE reduces to the MCE and hence the Guyer-Krumhansl-Stefan problem reduces to the Maxwell-Cattaneo-Stefan problem.

\section{Numerical Scheme}
In this section we consider the problem in the alternative variables $T(x,t)=u(\xi,t)$, $T_t(x,t)=v(\xi,t)$ and $q(x,t)=w(\xi,t)$, where $\xi=x/s$. The relevant transformed equations have been presented in the main text and take the form
\begin{subequations}\label{supp:num:system1}
\begin{alignat}{3}
        &\gamma v_t -\gamma\frac{s_t}{s}\xi v_\xi + v -\eta^2\frac{1}{s^2}v_{\xi\xi} -\frac{1}{s^2}u_{\xi\xi}=0,\qquad &&0<\xi<1,\\
        &v-u_t+\frac{s_t}{s}\xi u_\xi=0,\qquad &&0<\xi<1,\\
        &u=-1,\quad v=0,\qquad &\text{at }&\xi=0,\\
        &u=0,\qquad v+\frac{s_t}{s}u_\xi=0,\qquad &\text{at }&\xi=1,\\
        &u=v=0,\qquad &\text{at }&t=0,
\end{alignat}
\end{subequations}
and 
\begin{subequations}\label{supp:num:system2}
\begin{alignat}{3}
    &\gamma w_t-\gamma\frac{s_t}{s}\xi w_\xi+w- \eta^2\frac{1}{s^2} w_{\xi\xi}=-\frac{1}{s}u_{\xi},\qquad &&0<\xi<1,\label{supp:num:Q_fixed}\\
        &w_\xi=0,\qquad &\text{at }&\xi=0,\\
        &w_\xi=s_tu_\xi,\qquad &\text{at }&\xi=1,\\
        &w=0,\qquad &\text{at }&t=0.
\end{alignat}
\end{subequations}
The system \eqref{supp:num:system1} determines $u$ and $v$, whereas \eqref{supp:num:system2} is solved to determine $w$. Finally, the solid-liquid interface is determined by
\begin{equation}\label{supp:num:Stefan}
	\beta s_t=-w,\qquad \xi=1,
\end{equation}
subject to $s(0)=1$.

We discretize the unit interval into $N+1$ equidistant points of the form $\xi_i=i\Delta\xi$, where $i=0,...,N$ and $\Delta\xi=1/N$. Similarly, we consider time instants of the form $t_n=t_0+n\Delta t$, where $t_0$ is the starting time and $n\geq0$.

At each time step we first solve \eqref{supp:num:system1}, then use this information to update the heat flux by solving \eqref{supp:num:system2}, and finally we update the position of the solid-liquid interface using \eqref{supp:num:Stefan}. For this we discretize the derivatives of $u,v,w$ implicitly, i.e.
\begin{equation}
    u_t\approx\frac{u^{n+1}_i-u^n_i}{\Delta t},\qquad u_\xi\approx\frac{u^{n+1}_{i+1}-u^{n+1}_{i-1}}{2\Delta\xi},\qquad u_{\xi\xi}\approx\frac{u^{n+1}_{i+1}-2u^{n+1}_{i}+u^{n+1}_{i-1}}{(\Delta\xi)^2}
\end{equation}
whereas at $\xi=1$ we use 
\begin{equation}
    u_\xi|_{\xi=1}\approx\frac{3u^{n+1}_{N} -2u^{n+1}_{N-1} +u^{n+1}_{N-2}}{2\Delta\xi},
\end{equation}
to retain second-order accuracy in space. The boundary conditions for $w$ are discretized analogously. Upon writing $s\approx s^n$ and $s_t\approx s_t^n$, the discretized form of  \eqref{supp:num:system1} is a set of $2N+2$ linear equations that can be expressed as
\begin{equation}
    A^{n}_1\binom{\vec u^{n+1}}{\vec v^{n+1}}=A^{n}_2\binom{\vec u^{n}}{\vec v^{n}}-\vec e_1,
\end{equation}
for certain matrices $A^n_1,A^n_2\in\mathbb{R}^{2(N+1)\times2(N+1)}$ and where $\vec e_1=(1,0,....,0)^T\in\mathbb{R}^{2(N+1)}$ and $\vec u^{n+1}=(u^{n+1}_0,...,u^{n+1}_N)^T$, $\vec v^{n+1}=(v^{n+1}_0,...,v^{n+1}_N)^T$. To find $\vec w^{n+1}$ we discretize \eqref{supp:num:system2} at each grid point, using central differences in interior points and forwards/backwards differences at the boundaries. This gives a linear system for $\vec w^{n+1}$ of the form
\begin{equation}
    B^{n}_1\vec w^{n+1}=B^{n}_2\vec w^{n}-B^{n}_3\vec u^{n+1},
\end{equation}
for certain matrices $B^{n}_1,B^{n}_2,B^{n}_3\in\mathbb{R}^{(N+1)\times(N+1)}$. Finally, we obtain the updated value of the position and speed of the interface by
\begin{equation}
    s_t^{n+1}=-\beta^{-1}w^{n+1}_N,\qquad s^{n+1} = s^{n}+\Delta t s_t^{n+1}.
\end{equation}

\section{Asymptotic analysis of the Maxwell-Cattaneo-Stefan problem} \label{sec:maxwellcattaneo}
In this section we consider \eqref{supp:eqs} with $\eta=0$ and perform an asymptotic analysis of the problem assuming $\gamma,\beta\gg1$, which corresponds to a limit where memory effects are strong.  In the case when $\gamma = O(1)$ and $\beta\gg1$, we find that the leading-order solidification kinetics are identical to those predicted by Fourier's law.

We can combine Eqs. \eqref{supp:gk} and \eqref{supp:ce} to eliminate $Q$, which yields the hyperbolic heat equation (HHE)
\begin{equation}
    \gamma T_{tt}+T_{t}=T_{xx}.
\end{equation}
Mathematically, this equation describes the propagation of a temperature wave with a finite speed $\gamma^{-1/2}$. Damping is initially negligible, but appears gradually as the wave travels through the seed crystal. From a microscopic point of view, damping appears as the number of resistive collisions increases. 

Heat conduction through the crystal can be split into three time regimes. The first time regime captures the propagation of the disturbance at $x=0$ and the second time regime captures the collision of the wave against the solid-liquid interface. In the third time regime we enter a pseudo-steady state where the flux becomes a function of time and the solidification process eventually begins.

\subsection{First time regime} 
Due to the fact that $T=-1$ at $x=0$ and $s(0)=1$ we have $T,s=O(1)$. Additionally, the initial wave-like behaviour of the temperature implies $x=O(\gamma^{-1/2}t)$ and damping effects are negligible for $t\ll\gamma$. Finally, balancing terms in Eq. \eqref{supp:ce} requires $Q=O(\gamma^{-1/2})$. By introducing the scaled variables defined by $t=\hat t$, $x=\gamma^{-1/2}\hat x$, $Q=\gamma^{-1/2}\hat Q$, $T=\hat T$ and $s=\hat s$, the Stefan condition becomes $\hat s_{\hat t}=O(\gamma^{-1/2}\beta^{-1})$, hence $\hat s=1$ at leading order, which reduces the problem to
\begin{subequations}\label{maxcat:gammalarge_1}
\begin{alignat}{3}
    &{\hat Q}_{\hat t}+\gamma^{-1}\hat Q=-{\hat T}_{\hat x},\qquad &&\hat x,\hat t>0,\label{maxcat:gammalarge_1_mc}\\
    &{\hat T}_{\hat t}+{\hat Q}_{\hat x}=0,\qquad &&\hat x,\hat t>0,\\
    &\hat T=-1,\qquad &\text{at }&\hat x=0,\\
    &\hat T\to0,\qquad &\text{as }&\hat x\to\infty,\\
    &\hat T=\hat Q=0,\qquad &\text{at }&\hat t=0.
\end{alignat}
\end{subequations}
The leading-order problem can be solved using the method of characteristics, giving $\hat T=-\mathcal{H}(\hat t-\hat x)$ and $\hat Q=-\mathcal{H}(\hat t-\hat x)$, where $\mathcal{H}$ is the Heaviside function.  The solution therefore corresponds to a travelling wave that propagates with constant velocity.  The position of the wavefront is given by $\hat{x}_f(\hat{t}) = \hat{t}$.  An examination of the large-time behaviour shows that $\hat{x}_f = O(\hat{t})$ and $\hat Q,\hat T=O(1)$ as $\hat{t} \to \infty$, which determine the scales for the second time regime.

\subsection{Second time regime}\label{sec:maxcat:gammalarge_1}
By looking at the scaled system \eqref{maxcat:gammalarge_1}, a new balance arises for $\hat t=O(\gamma)$, which implies $x=O(\gamma^{1/2})$ and hence $x\gg1$. This suggests that one should first consider the time regime where the heat wave hits the boundary, i.e., we force $x_f =O(1)$ and hence $t=O(\gamma^{1/2})$. Additionally, we have $T,s=O(1)$ and $Q=O(\gamma^{-1/2})$ in order to match to the solution in the previous time regime. In the corresponding scaled variables $\tilde x,\tilde s,\tilde t,\tilde Q$ and $\tilde T$, the Stefan condition becomes ${\tilde s}_{\tilde t}=O(\beta^{-1})$, hence we still do not have any interface motion at leading order. In these variables, the problem is similar to \eqref{maxcat:gammalarge_1}, with the only difference in the spatial domain,
\begin{subequations}\label{maxcat:gammalarge_2}
\begin{alignat}{3}
    &{\tilde Q}_{\tilde t}+\gamma^{-1/2}\tilde Q=-{\tilde T}_{\tilde x},\qquad &&0<\hat x<1,\label{maxcat:gammalarge_2_mc}\\
    &{\tilde T}_{\tilde t}+{\tilde Q}_{\tilde x}=0,\qquad &&0<\tilde x<1,\label{maxcat:gammalarge_2_ce}\\
    &\tilde T=-1,\qquad &\text{at }&\tilde x=0,\\
    &\tilde T=0,\qquad &\text{at }&\tilde x=1,
\end{alignat}
\end{subequations}
with $\tilde{T} \sim -\mathcal{H}(\tilde{t} - \tilde{x})$ and $\tilde{Q} \sim -\mathcal{H}(\tilde{t} - \tilde{x})$ as $\tilde{t} \sim 0$ as matching conditions.

We first solve the leading-order problem for the temperature, which satisfies the wave equation
$\tilde{T}_{\tilde{t} \tilde{t}} = \tilde{T}_{\tilde{x} \tilde{x}}$ in the bulk.  The matching condition corresponds to a right-moving travelling wave and satisfies the bulk equation. Thus, the solution for $0 < \tilde{t} < 1$ is given by $\tilde{T} = -\mathcal{H}(\tilde{t} - \tilde{x})$.  When $\tilde{t} = 1$, this wave collides with the boundary at $\tilde{x} = 1$ and reflects.  To determine the dynamics of this reflection, we shift the origin using the transformation $\tilde{x} = \tilde{x}' + 1$ and write the solution for the temperature in terms of an incoming wave $\tilde{T}_I$ and a reflected wave $\tilde{T}_R$ as $\tilde{T} = \tilde{T}_I(\tilde{t} - \tilde{x}') + T_R(\tilde{t} + \tilde{x}')$.  Imposing the condition $\tilde{T} = 0$ when $\tilde{x}' = 0$ gives that $T_R = -T_I$, implying that the reflected wave destructively interferes with the incoming wave.  The resulting solution for the temperature is a left-moving travelling wave given by $\tilde{T} = -\mathcal{H}(\tilde{t} + \tilde{x} - 2)$ valid for $1 < \tilde{t} < 2$.  When $\tilde{t} = 2$, the wave collides with the left boundary at $\tilde{x} = 0$, again creating a reflected wave.  By writing the temperature as $\tilde{T} = -1 + \tilde{T}'$, the same analysis can be used to show that the reflected wave destructively interferes with the incoming wave, resulting in a right-moving travelling wave given by $\tilde{T} = -\mathcal{H}(\tilde{t} - \tilde{x} - 2)$ for $2 < \tilde{t} < 3$.  As this solution is identical (up to a translation in time) to that for $0 < \tilde{t} < 1$, the sequence of destructive reflections repeats itself.  Importantly, we find that $\tilde{T} = O(1)$ for all times in this regime. 

The dynamics for the flux are much more interesting and can be determined by solving \eqref{maxcat:gammalarge_2_mc} using the known solutions for the temperature.  For $0 < \tilde{t} < 1$, the flux is also a right-moving travelling wave given by $\tilde{Q} = -\mathcal{H}(\tilde{t} - \tilde{x})$.  However, the collision with the boundary at $\tilde{x} = 1$ results in constructive interference.  This can be seen by writing the solution for the temperature for $1 < \tilde{t} < 2$ as $\tilde{T} = \tilde{T}_I(\tilde{t} - \tilde{x}') - \tilde{T}_I(\tilde{t} + \tilde{x}')$.  The corresponding solution for the flux is $\tilde{Q} = \tilde{T}_I(\tilde{t} - \tilde{x}') + \tilde{T}_I(\tilde{t} + \tilde{x}')$; thus, the right-moving incoming wave and left-moving reflected wave add in a constructive manner.  We find that $\tilde{Q} = -1 - \mathcal{H}(\tilde{t} + \tilde{x} - 2)$ for $1 < \tilde{t} < 2$, corresponding to a left-moving travelling wave which takes on values between $-2$ and $-1$.  Repeating the analysis shows that the collision with the boundary at $\tilde{x} = 0$ is also constructive, resulting in a solution for the flux given by $\tilde{Q} = -2 - \mathcal{H}(\tilde{t} - \tilde{x} - 2)$ for $2 < \tilde{t} < 3$, which, again, is a right-moving travelling wave.  In this case, the flux ranges from $-3$ to $-2$ across $\tilde{x}$. The sequence of constructive reflections continues, each time reducing the flux by one unit.  Thus, the asymptotic behaviour of the flux is given by $\tilde{Q} = O(\tilde{t})$ for $\tilde{t} \gg 1$.  

Having determined the behaviour of the temperature and flux for large times, we are now in a position to determine the relevant scales for the next time regime.  It turns out there are two main choices depending on the relative sizes of $\beta$ and $\gamma$.  One one hand, if $\gamma\ll\beta$, a new balance in the governing equations is obtained by setting $\tilde t=O(\gamma^{1/2})$ and $\tilde{Q} = O(\gamma^{1/2})$,
\begin{align}
    {\tilde Q}_{\tilde t}+\tilde Q &= -{\tilde T}_{\tilde x},\\
    {\tilde Q}_{\tilde x} &= O(\gamma^{-1}),
\end{align}
whereas the Stefan condition becomes
\begin{equation}
    {\tilde s}_{\tilde t} = O(\gamma\beta^{-1}).
\end{equation}
This scaling implies that the travelling waves start to become damped, which is associated with a relaxation to the classical Fourier profiles; however, solidification has yet to begin. On the other hand, if $\gamma\gg\beta$, then we can balance both sides of the Stefan condition by setting $\tilde t=O(\beta^{1/2})$ and $\tilde{Q} = O(\beta^{1/2})$. In this case, the rescaled equations are
\begin{align}
    \tilde{Q}_{\tilde t} &= -\tilde{T}_{\tilde x} + O(\gamma^{1/2}\beta^{-1/2}),\\
    \tilde{Q}_{\tilde x} &= O(\beta^{-1}), \\
    \tilde{s}_{\tilde{t}} &= -\tilde{Q},
\end{align}
implying that solidification begins before the thermal waves become damped.  Both of these scalings result in a quasi-steady scenario whereby the flux is uniform ($\tilde{Q}_x = 0$) and the temperature profile is linear (which is seen by differentiating the MCEs with respect to $\tilde{x}$).

\subsection{Third time regime} \label{sec:maxwellcattaneo_third}
As previously discussed, the choice of the time scale depends on the relative size of $\gamma$ to $\beta$. For simplicity, we can consider the distinguished limit where $\gamma^{-1/2}\beta^{1/2}=\alpha=O(1)$. Based on the large-time analysis of the previous time regime we take $t=O(\gamma^{1/2}\beta^{1/2})$, $Q=O(\gamma^{-1/2}\beta^{1/2})$ and $x,s,T=O(1)$. In the corresponding scaled variables $\bar t,\bar x,\bar s,\bar Q$ and $\bar T$, the problem becomes
\begin{subequations}\label{maxcat:gammalarge_3}
\begin{alignat}{3}
    &{\bar Q}_{\bar t}+\alpha\bar Q=-{\bar T}_{\bar x},\qquad &&0<\bar x<\bar s,\label{maxcat:gammalarge_3_mc}\\
    &{\bar Q}_{\bar x}=0,\qquad &&0<\bar x<\bar s,\label{maxcat:gammalarge_3_ce}\\
    &\bar T=-1,\qquad &\text{at }&\bar x=0,\label{maxcat:gammalarge_3_bc0}\\
    &\bar T=0,\qquad &\text{at }&\bar x=\bar s,\label{maxcat:gammalarge_3_bcs}\\
    &{\bar s}_{\bar t}=-\bar Q,\qquad &\text{at }&\bar x=\bar s,\label{maxcat:gammalarge_3_Stefan}
\end{alignat}
\end{subequations}
with matching conditions given by the previous time regime. Differentiating Eq. \eqref{maxcat:gammalarge_3_mc} with respect to $x$ gives that $T$ is linear in space and hence, applying Eqs. \eqref{maxcat:gammalarge_3_bc0} and \eqref{maxcat:gammalarge_3_bcs}, we find
\begin{equation}
    \bar T(\bar t,\bar x)=\frac{\bar x}{\bar s}-1.
\end{equation}
Using this expression in \eqref{maxcat:gammalarge_3_mc} yields
\begin{equation}
  {\bar Q}_{\bar t}+\alpha\bar Q=-\frac{1}{\bar s},
  \label{maxcat:gammalarge_3_Qode}
\end{equation}
subject to the initial condition $\bar Q(0)=0$ due to the large time behaviour $\tilde Q\sim-\tilde t$ in the previous time regime. Note, for $\bar{t}\gg1$ we find $\bar Q \sim -1/(\alpha\bar s)$ and hence, using \eqref{maxcat:gammalarge_3_Stefan}, the classical behaviour $\bar s\sim\bar t^{1/2}$ is obtained. The reduced model given by \eqref{maxcat:gammalarge_3_Stefan} and \eqref{maxcat:gammalarge_3_Qode} can also be used when $\alpha \ll 1$, corresponding to the case when $\gamma \gg \beta$. From an asymptotic point of view, this case can be studied using matched asymptotic expansions in terms of $\alpha$ through the introduction of additional time and length scales.

\section{Asymptotic analysis of the Guyer-Krumhansl-Stefan problem}\label{sec:gk}
As we show now, the presence of non-local effects leads to a completely different small-time behaviour of the problem. This is due to the fact that governing equations are mathematically very different in the cases $\eta=0$ and $\eta>0$. In the first case we obtain the HHE, which is a hyperbolic equation and describes the propagation of a heat wave, contrary to the GKHE, which is a parabolic equation.  In this section we will perform an asymptotic analysis of \eqref{supp:eqs} assuming that either memory or non-local effects are strong. 

\subsection{Enhanced memory effects}\label{sec:gk:memory}
Let us assume that $\gamma\gg\eta=O(1)$, i.e. that heat transport in the seed crystal is initially dominated by memory effects. Additionally we assume $\beta\gg1$.

The initial heat conduction through the solid can be divided into four time regimes. In the first time regime, heat conduction is described by an effective Fourier law. In the second time regime, both memory and non-local effects enter the leading-order problem, giving rise to a wave-like propagation of heat that is analogous to one-dimensional viscoelastic wave propagation. The third time regime captures the refections of the thermal waves with the boundaries. In the fourth regime, the system tends towards a quasi-steady state and solidification begins. Non-local effects are negligible but memory effects persist.

\subsubsection{First time regime} 
At the beginning of the process we have $T,s=O(1)$ due to the boundary condition at $x=0$ and the initial condition for $s$. Assume now $t=O(\epsilon)$, where $\epsilon$ is an arbitrarily small parameter. Balancing terms in \eqref{supp:ce} requires $xQ^{-1}=O(\epsilon)$. Since $\epsilon\ll1$ and $\gamma\gg1$, the terms $Q_t$ and $Q$ in \eqref{supp:gk} cannot balance. The same happens for the terms $T_x$ and $Q_{xx}$ on the r.h.s. of \eqref{supp:gk}. Hence, $Q_t$ must balance either $T_x$ or $Q_{xx}$. The first case yields $x=O(\gamma^{-1}\epsilon^2)$ and $Q=O(\gamma^{-1}\epsilon)$, which would eliminate $Q_t$ from the leading order equation. Hence, the only sensible balance is given by the two non-classical contributions, which gives $x=O(\gamma^{-1/2}\epsilon^{1/2})$ and $Q=O(\gamma^{-1/2}\epsilon^{-1/2})$.
In terms of the new variables defined by $T=\hat T$, $x=\gamma^{-1/2}\epsilon^{1/2}\hat x$, $s=\hat s$, $t=\epsilon\hat t$ and $Q=\gamma^{-1/2}\epsilon^{-1/2}\hat Q$, the Stefan condition becomes ${\hat s}_{\hat t}=O(\gamma^{-1/2}\beta^{-1}\epsilon^{1/2})$, thus $\hat s\approx1$, and the G-K equation becomes
\begin{equation}\label{supp:gk_large_gamma_first}
	{\hat Q}_{\hat t}+\gamma^{-1}\epsilon \hat Q=-\epsilon{\hat T}_{\hat x}+\eta^2{\hat Q}_{\hat x\hat x}.
\end{equation}
Since $\epsilon$ is arbitrarily small, the leading order problem becomes
\begin{subequations}
\begin{alignat}{3}
    &{\hat Q}_{\hat t}=\eta^2{\hat Q}_{\hat x\hat x},\qquad &&\hat x,\hat t>0,\label{gk:gammalarge_1_gk}\\
	&{\hat T}_{\hat t}+{\hat Q}_{\hat x}=0,\qquad &&\hat x,\hat t>0,\label{gk:gammalarge_1_ce}\\
	&\hat T = -1,\qquad &\text{at }&\hat x=0,\\
	&\hat T = 0,\qquad &\text{as }&\hat x\to\infty,\\
	&\hat q=\hat T=0,\qquad &\text{at }&\hat t=0,\label{gk:gammalarge_1_ic}
\end{alignat}
\end{subequations}
Equations \eqref{gk:gammalarge_1_gk} and \eqref{gk:gammalarge_1_ce} can be combined to give
\begin{equation}
	{\hat Q}_{\hat t}=-\eta^2{\hat T}_{\hat x\hat t},
\end{equation}
from where we recover Fourier's law after integrating with respect to time and applying Eq. \eqref{gk:gammalarge_1_ic}. The problem is therefore equivalent to a classical heat conduction problem in a semi-infinite domain with an effective thermal conductivity $\eta^2$. As suggested by the fact that the length scale has remained undetermined, this problem admits a solution in terms of a similarity variable $\hat y=\hat x/\sqrt{\hat t}$. In terms of the original non-dimensional variables, the solution takes the form
\begin{equation}\label{gk:gammalarge_1_sol}
	T(x,t)=\erf\left(\frac{x}{2\sqrt{\zeta t}}\right)-1,\qquad Q(x,t)=-\sqrt{\frac{\zeta}{\pi t}}\exp\left(-\frac{x^2}{4\zeta t}\right),
\end{equation}
where $\zeta=\eta^{2}/\gamma$.

\subsubsection{Second time regime} 
Equation \eqref{supp:gk_large_gamma_first} indicates that the previous time regime breaks down for $t=O(1)$, since the term $T_x$ enters the leading order problem. From the previous time regime we find $x,Q=O(\gamma^{-1/2})$, whereas $s,T=O(1)$. Upon defining the new variables $T=\tilde T$, $x=\gamma^{-1/2}\tilde x$, $s=\tilde s$, $t=\tilde t$ $q=\gamma^{-1/2}\tilde q$, we obtain again $\tilde s_{\tilde t}=O(\gamma^{-1/2}\beta^{-1})$, hence $\tilde s\approx 1$, and
\begin{subequations}\label{gk:gammalarge_2}
\begin{alignat}{3}
	&{\tilde T}_{\tilde t}+{\tilde Q}_{\tilde x}=0,\qquad &&\tilde x,\tilde t>0,\label{gk:gammalarge_2_ce}\\
	&{\tilde Q}_{\tilde t}=-{\tilde T}_{\tilde x}+\eta^2{\tilde Q}_{\tilde x\tilde x},\qquad &&\hat x,\hat t>0,\label{gk:gammalarge_2_gk}\\
	&\tilde T = -1,\qquad &\text{at }&\tilde x=0,\label{gk:gammalarge_2_bc0}\\
	&\tilde T = 0,\qquad &\text{as }&\tilde x\to\infty.
\end{alignat}
\end{subequations}
By eliminating the flux $\tilde{Q}$, we find that the temperature satisfies a linear viscoelastic equation given by
\begin{align}
    {\tilde T}_{\tilde t\tilde t}={\tilde T}_{\tilde x\tilde x}+\eta^2{\tilde T}_{\tilde x\tilde x\tilde t},
    \label{gk:gammalarge_2_T2}
\end{align}
with the additional condition that $\tilde{T}_{\tilde{t}}(\tilde{x},0) = 0$.  The large-time limit of \eqref{gk:gammalarge_2_T2} implies that the solution for the temperature is given by a right-moving travelling wave that has an interior, moving boundary layer located at $\tilde{x} = \tilde{t}$ that captures weak non-local (``viscous'') effects.  The solution for the temperature for $\tilde{x} = O(\tilde{t})$ with $\tilde{t} \gg 1$ can be written as
\begin{align}
\tilde{T}(\tilde{x},\tilde{t}) = 
\begin{cases}
\tilde{T}_\text{up}(\tilde{x} - \tilde{t}), &\quad \tilde{x} < \tilde{t}, \\
\tilde{T}_\text{inner}(\tilde{x},\tilde{t}), &\quad \tilde{x} - \tilde{t} = O(1), \\
\tilde{T}_\text{down}(\tilde{x} - \tilde{t}), &\quad \tilde{t} < \tilde{x},
\end{cases}
\end{align}
where the upstream and downstream solutions $\tilde{T}_\text{up}$ and $\tilde{T}_\text{down}$ are, in principle, obtained by matching to the solution for $O(1)$ times and the inner solution in the boundary layer is
\begin{align}
\tilde{T}_\text{inner} = \frac{1}{2}(\tilde{T}_\text{up}(0) + \tilde{T}_\text{down}(0)) + \frac{1}{2}(\tilde{T}_\text{down}(0) - \tilde{T}_\text{up}(0))\erf\left(\frac{\tilde{x} - \tilde{t}}{2^{1/2} \eta \tilde{t}^{1/2}}\right).
\end{align}
This solution reveals that the temperature will remain $O(1)$ in size for large times.  The large-time approximation to the flux can be obtained via \eqref{gk:gammalarge_2_ce} and shown to remain $O(1)$ in size for large times as well.  The boundary layer at $\tilde{x} = \tilde{t}$ acts as the diffuse wavefront of the propagating thermal wave.

\subsubsection{Third time regime} 
This time regime is analogous to that analysed in Sec.~\ref{sec:maxcat:gammalarge_1} for the MCE, in which thermal waves propagate and collide with the boundaries of the crystal.  The first collision occurs at the right boundary when $t = \gamma^{1/2}$.  Therefore, in this time regime we choose a length scale of $x = O(1)$ and a time scale of $t = O(\gamma^{1/2})$.  The heat flux has not changed in magnitude, therefore we choose a flux scale of $Q=O(\gamma^{-1/2})$ as in the previous time regime. Additionally, we have $s,T=O(1)$. In terms of the corresponding scaled variables $\bar x,\bar s,\bar t,\bar Q$ and $\bar T$, the Stefan condition takes the form $\bar s_{\bar t}=O(\beta^{-1})$ and hence $\bar s\approx1$, which leads to the problem
\begin{subequations}
\begin{alignat}{3}
    &\bar{T}_{\bar{t}\bar{t}} + \gamma^{-1/2}\bar{T}_{\bar{t}} = \bar{T}_{\bar{x}\bar{x}} + \eta^2 \gamma^{-1/2} \bar{T}_{\bar{x}\bar{x}\bar{t}},    \qquad &&0<\bar x<1,
    \label{gk:gammalarge_3_gk}\\
    &{\bar T}_{\bar t}+{\bar Q}_{\bar x}=0,\qquad &&0<\bar x<1,\\
	&\bar T = -1,\qquad &\text{at }&\bar x=0,\\
	&\bar T = 0,\qquad &\text{at }&\bar x=\bar 1,
\end{alignat}
\end{subequations}
where the flux has been eliminated from the GKE to produce \eqref{gk:gammalarge_3_gk}.  Naively taking $\gamma \to \infty$ in the GKHE \eqref{gk:gammalarge_3_gk} shows that the temperature satisfies the same leading-order problem as in Sec.~\ref{sec:maxcat:gammalarge_1} and thus we can expect similar wave propagation and destructive interference to occur.  The main difference is that non-local effects will influence the temperature and flux profiles near the wavefront. Remarkably, they do not influence the reflection dynamics to leading order.

We first consider the dynamics that occur before the thermal wave collides with the right boundary ($0 < \bar{t} < 1$).  The leading-order solutions for the temperature and flux can be constructed using matched asymptotic expansions and are found to be given by $\bar{T}(\bar{x},\bar{t}) = \mathcal{F}(\bar{x},\bar{t})$ and $\bar{Q}(\bar{x},\bar{t}) = \mathcal{F}(\bar{x},\bar{t})$ where
\begin{align}
\mathcal{F}(\bar{x},\bar{t}) \equiv \frac{1}{2}\,\erf\left(\frac{\gamma^{1/4}(\bar{x} - \bar{t})}{2^{1/2} \eta \bar{t}^{1/2}}\right) -\frac{1}{2}.
\label{gk:gammalarge_3_Tsol1}
\end{align}
Equation \eqref{gk:gammalarge_3_Tsol1} is analogous to the Heaviside function that was found in Sec.~\ref{sec:maxcat:gammalarge_1} but it now accounts for the diffuse nature of the wave front which has a finite width of $O(\gamma^{-1/4} \eta t^{1/2})$. 

To understand the dynamics of the reflection that occurs at $\bar{x} = 1$ when $\bar{t} = 1$, we write $\bar{x} = 1 + \eta^2\gamma^{-1/2} \bar{x}'$ and $\bar{t} = 1 + \eta^2\gamma^{-1/2} \bar{t}'$ so that the temperature now satisfies $\bar{T}_{\bar{t}'\bar{t'}} = \bar{T}_{\bar{x}'\bar{x}'} + \bar{T}_{\bar{x}'\bar{x}'\bar{t'}}$.  The bulk equation admits wave solutions of the form $\bar{T} \sim \e^{i(k \bar{x}' - \omega \bar{t}')}$, where $\omega = \pm (ik/2)[k + (k^2 - 4)^{1/2}]$.  Since $\omega$ is complex for wavenumbers that satisfy $k < 2$, these short-wave modes are attenuated during wave propagation.  By considering a single wave mode, writing the temperature as an incoming and reflected wave, $\bar{T} = \bar{T}_I \e^{i(k\bar{x}'-\omega \bar{t}')} + \bar{T}_R\e^{i(-kx - \omega \bar{t}')}$, and applying the boundary condition $\bar{T} = 0$ when $\bar{x}'=0$, we find that $\bar{T}_R = -\bar{T}_I$.  That is, the reflected wave has the same amplitude as the incoming wave but opposite sign, leading to destructive interference again.  Moreover, non-local effects to not alter the amplitude of the reflected wave.  By considering a single-mode wave solution for the flux, we find from conservation of energy that $\bar{Q} = (\omega/k)[\bar{T}_I\e^{i(k\bar{x}'-\omega \bar{t}')} + \bar{T}_I\e^{i(-k\bar{x}'-\omega\bar{t}')}]$.  Therefore, the reflection of the flux wave leads to the same constructive interference as before.

Using the superposition principle, the solutions for the temperature and flux after the first reflection, but before the second reflection with the left boundary, can be written as
\begin{subequations}
    \label{gk:gammalarge_3_Tsol2}
\begin{align}
    \bar{T} = \mathcal{F}(\bar{x}, \bar{t}) - \mathcal{F}(2-\bar{x}, \bar{t}), \\
    \bar{Q} = \mathcal{F}(\bar{x}, \bar{t}) + \mathcal{F}(2-\bar{x}, \bar{t}),
\end{align}
\end{subequations}
for $1 < \bar{t} < 2$.  In fact, since the second terms of \eqref{gk:gammalarge_3_Tsol2} are exponentially small for $\bar{t} < 1$, these expressions are valid for $0 < \bar{t} < 2$.  By further exploiting the asympototic limits of the function $\mathcal{F}$, the solution for times given by $0 < \bar{t} < 2N$, where $N$ is an integer, is given by
\begin{subequations}
\label{gk:gammalarge_3_soln}
\begin{align}
\bar{T}(\bar{x},\bar{t}) = \sum_{n=0}^{N-1} \left[\mathcal{F}(\bar{x} + 2n, \bar{t}) - \mathcal{F}(2(n+1)-\bar{x},\bar{t})\right], \\
\bar{Q}(\bar{x},\bar{t}) = \sum_{n=0}^{N-1} \left[\mathcal{F}(\bar{x} + 2n, \bar{t}) + \mathcal{F}(2(n+1)-\bar{x},\bar{t})\right].
\end{align}
\end{subequations}
Physically, the expression in \eqref{gk:gammalarge_3_soln} can be interpreted as a collection of right- and left- moving travelling waves, each of which has a diffuse wavefront that broaden in time according to a $O(t^{1/2})$ scaling law.  From these solutions we can deduce that $\bar{T} = O(1)$ and $\bar{Q} = O(\bar{t})$ as $\bar{t} \to \infty$. 

We are now in a position to determine the next time regime using the asymptotic behaviour of the temperature and flux for large $\bar{t}$.  Using the current scaling, the GKE can be written as $\bar{Q}_{\bar{t}} + \gamma^{-1/2} \bar{Q} = -\bar{T}_{\bar{x}} + \eta^2\gamma^{-1/2} \bar{Q}_{\bar{x}\bar{x}}$, indicating that the second term on the left-hand side becomes relevant when $\bar{t} = O(\gamma^{1/2})$.  Alternatively, balancing both terms in the Stefan condition implies that solidification will occur when $\bar{t} = O(\beta^{1/2})$.  The choice of the time scale will therefore depend on the relative size $\gamma$ to $\beta$.

\subsubsection{Fourth time regime} \label{sec:gk:memory_fourth}
We consider the distinguished limit where $\gamma^{-1/2}\beta^{1/2}=\alpha=O(1)$. We choose $t,Q=O(\gamma^{1/2}\beta^{1/2})$, whereas $x,s,T=O(1)$. In the new variables $\check x, \check s,\check t,\check Q$ and $\check T$, the leading order problem takes the form
\begin{subequations}
\begin{alignat}{3}
    &{\check Q}_{\check t}+\alpha\check Q=-{\check T}_{\check x},\qquad &&0<\check x<\check s,\label{gk:gammalarge_4_gk}\\
    &{\check Q}_{\check x}=0,\qquad &&0<\check x<\check s,\\
    &\bar T = -1,\qquad &\text{at }&\bar x=0,\\
	&\bar T = 0,\qquad &\text{at }&\bar x=\bar s,\\
	&{\check s}_{\check t}=-\check Q,\qquad &\text{at }&\bar x=\bar s, 
\end{alignat}
\end{subequations}
with matching conditions provided by the previous time regime. This system is analogous to \eqref{maxcat:gammalarge_3} and hence the GKE has effectively reduced to the MCE, since non-local effects have become negligible. Interestingly, this has happened before the solidification process has begun.

\begin{figure}[h!]
    \centering
    \begin{subfigure}{.47\textwidth}
	\includegraphics[width=\textwidth]{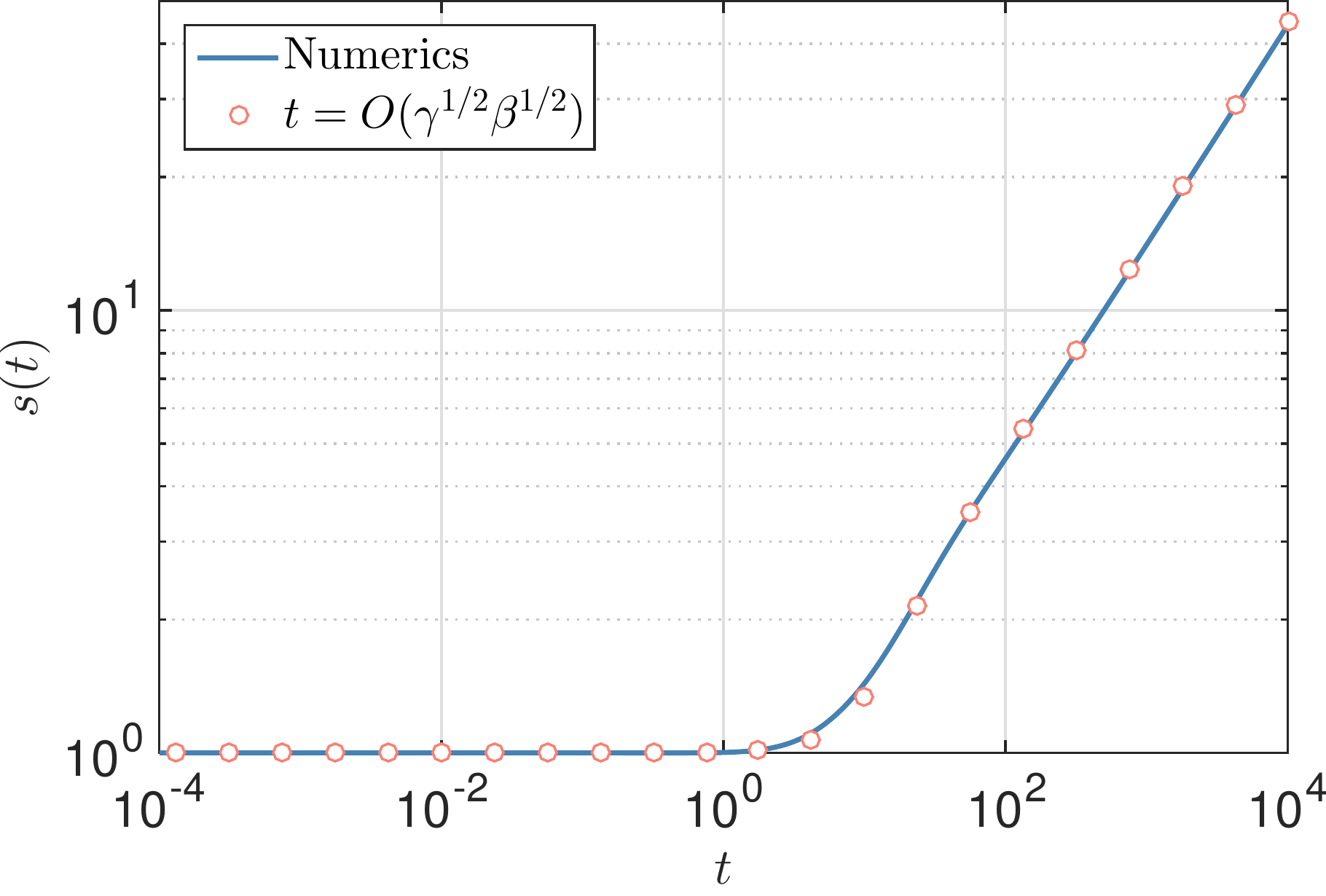}
    \caption{$\gamma=10,\beta=10$.}
	\end{subfigure}
    ~
    \begin{subfigure}{.47\textwidth}
	\includegraphics[width=\textwidth]{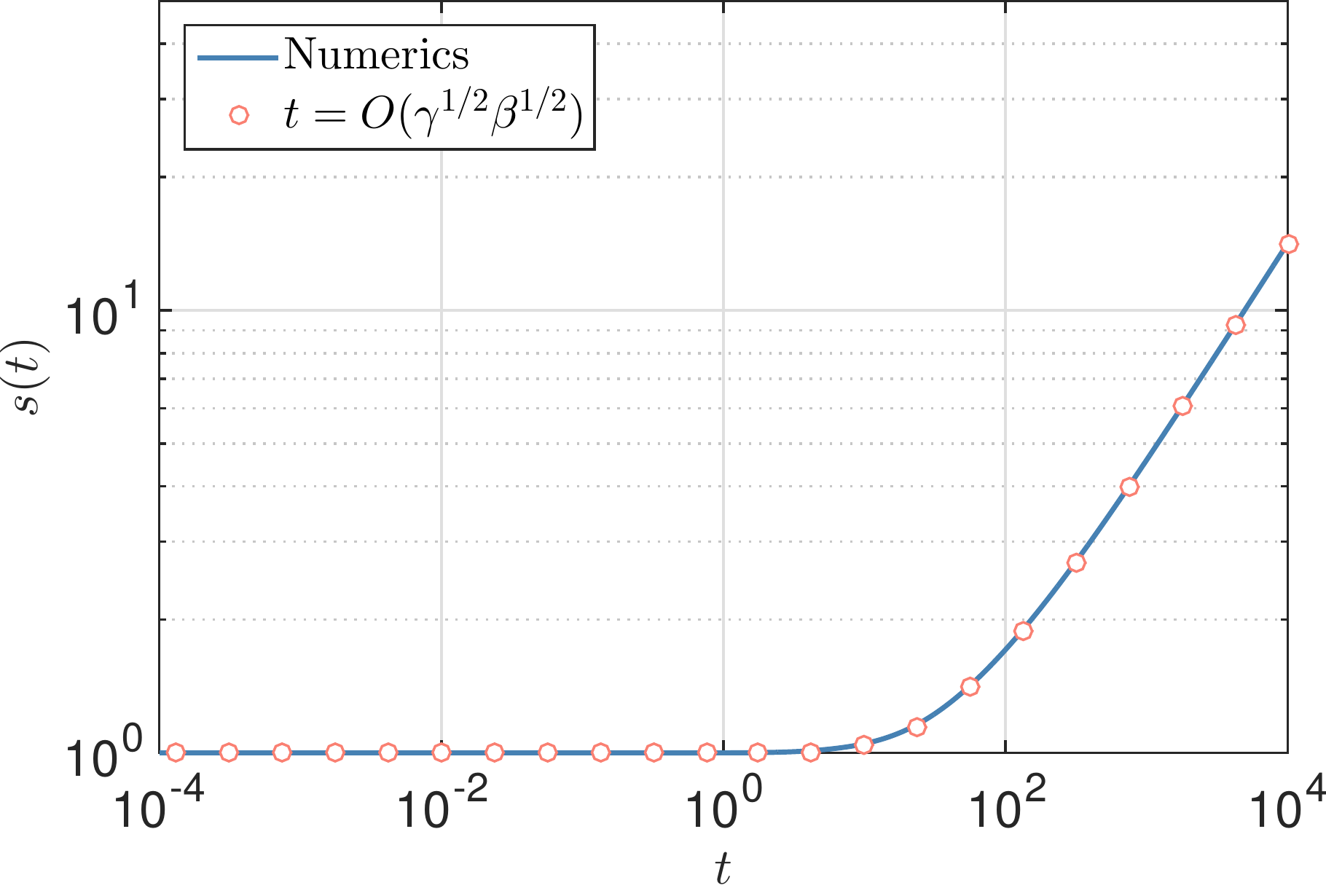}
    \caption{$\gamma=10$, $\beta=100$.}
	\end{subfigure}
	
    \begin{subfigure}{.47\textwidth}
	\includegraphics[width=\textwidth]{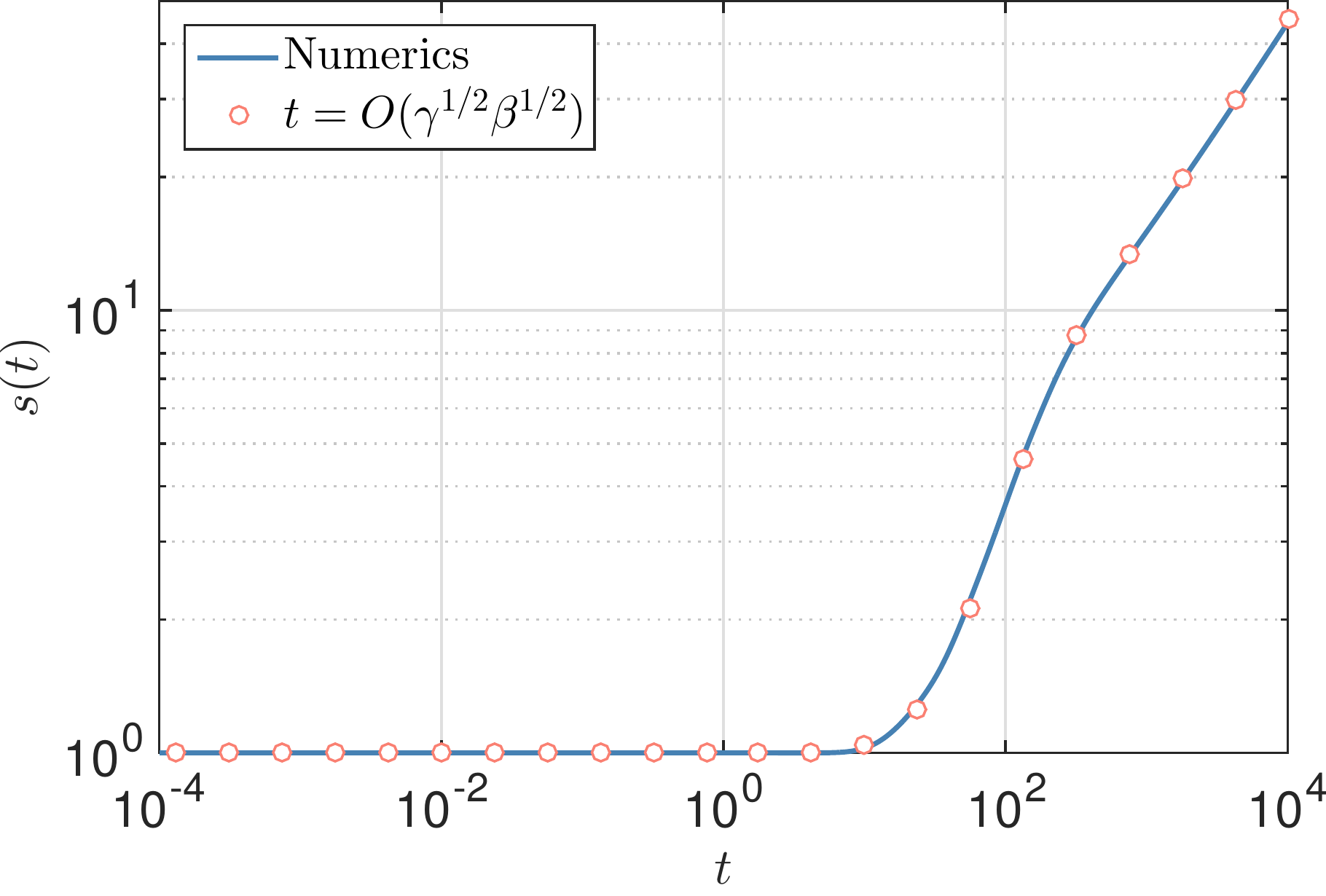}
    \caption{$\gamma=100,\beta=10$.}
	\end{subfigure}
    ~
    \begin{subfigure}{.47\textwidth}
	\includegraphics[width=\textwidth]{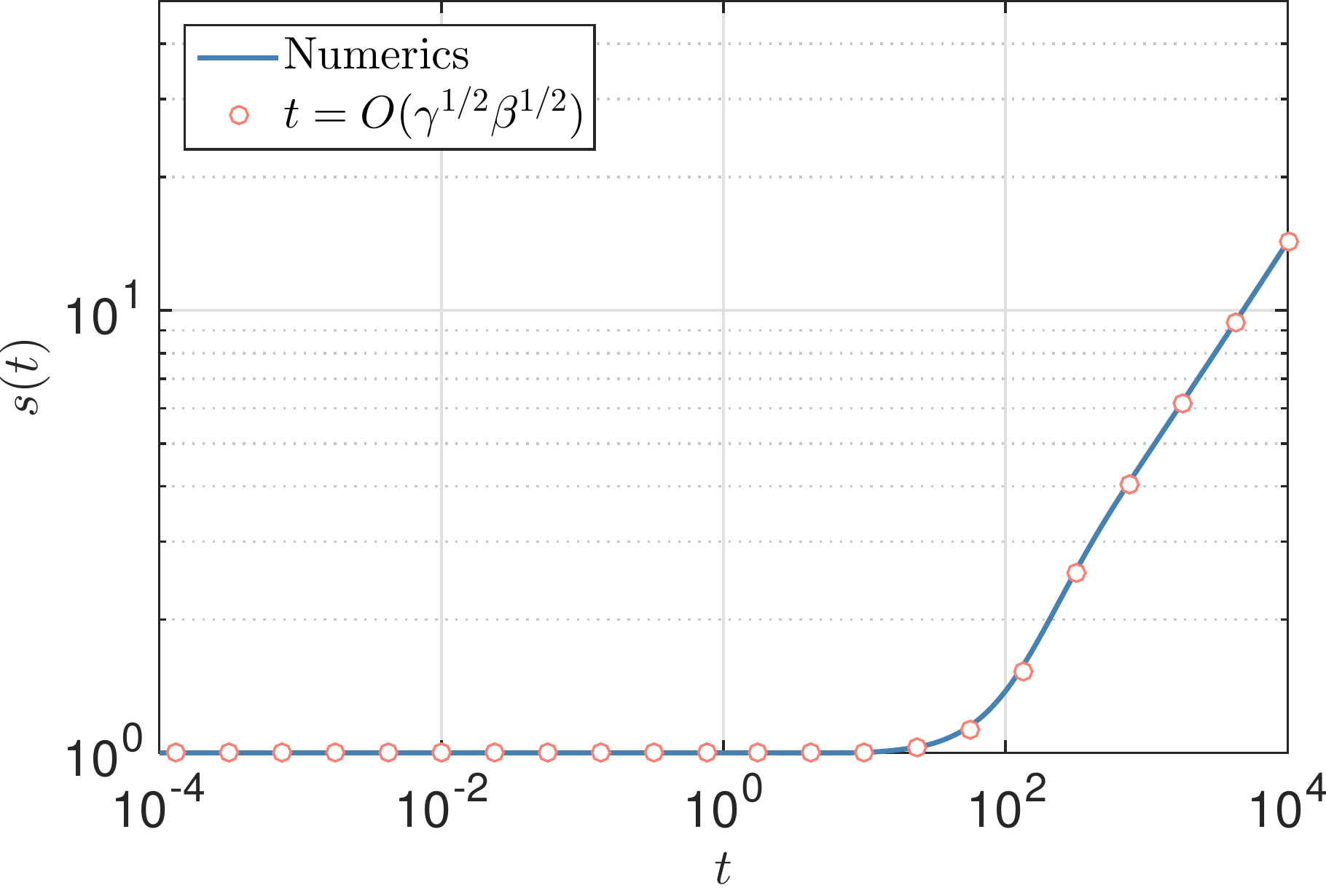}
    \caption{$\gamma=100$, $\beta=100$.}
	\end{subfigure}
    \caption{Evolution of the interface for different values of $\beta$ and $\gamma$, according to numerical simulation and the solution of the ODE in the fourth time regime. The dimensionless mean free path has been set to unity.}
    \label{fig:num_asy1}
\end{figure}

\subsection{Enhanced non-local effects}
Here we assume $\beta,\eta^2\gg\gamma=O(1)$, therefore non-local effects are the dominant non-classical feature of heat transport. There are three time regimes to consider. The first time regime captures the initial heat flux through the seed crystal described by an effective form of Fourier's law. In the second time regime, both memory and non-local terms enter the leading-order problem; however, unlike the case of enhanced memory effects where the temperature satisfied a viscoelastic wave equation, now the problem is quasi-steady and the temperature profile is linear. Solidification begins in the second time regime but eventually terminates due to the diminishing influence of the non-classical conduction mechanisms that drive phase change.  In the third time regime, the classical contribution to the flux from temperature gradients becomes dominant, allowing solidification to restart. 

\subsubsection{First time regime} 
We consider heat flow across the entire seed crystal and take $x = O(1)$. Since $T = O(1)$, conservation of energy implies that $Q = O(t)$.  The only sensible balance in Eq. \eqref{supp:gk} is obtained by balancing the non-classical contributions, which gives $t = O(\eta^{-2})$. Defining $T=\hat T$, $x=\hat x$, $s=\hat s$, $t=\eta^{-2}\hat t$ $Q=\eta^{2}\hat Q$ we find $\hat s_{\hat t}=O(\beta^{-1})$ and thus solidification is negligible at leading order. Upon neglecting terms of order $\eta^{-4}$, the remaining equations become
\begin{subequations}
\begin{alignat}{3}
	&\gamma{\hat Q}_{\hat t}+\eta^{-2}{\hat Q}={\hat Q}_{\hat x\hat x},\qquad &&0<\hat x<1,\,\hat t>0,\label{gk:etalarge_1_gk}\\
	&{\hat T}_{\hat t}+{\hat Q}_{\hat x}=0,\qquad &&0<\hat x<1,\,\hat t>0,\label{gk:etalarge_1_ce}\\
	&\hat T = -1,\qquad &\text{at }&\hat x=0,\\
	&\hat T = 0,\qquad &\text{at }&\hat x=1,\\
	&\hat Q=\hat T=0,\qquad &\text{at }&\hat t=0,\label{gk:etalarge_1_ic}
\end{alignat}
\end{subequations}
We can combine \eqref{gk:etalarge_1_ce} and \eqref{gk:etalarge_1_gk} to obtain (after neglecting the $O(\eta^{-2})$ term)
\begin{equation}
	\gamma{\hat Q}_{\hat t} = -{\hat T}_{\hat x\hat t},
\end{equation}
from where we recover Fourier's law. The problem is therefore equivalent to a classical heat conduction problem in the unit interval and with an effective thermal conductivity $\gamma^{-1}$. The large-time solution for the temperature is $\hat T\sim\hat x-1$, whereas $\hat Q\sim-\gamma^{-1}$.

There are two possible choices for the next time regime. On one hand, for $\hat t=O(\beta)$ the terms in the Stefan condition balance and the solidification enters the leading order problem. On the other hand, a new balance in \eqref{gk:etalarge_1_gk} is obtained for $\hat t=O(\eta^2)$. The choice of the next time scale therefore depends on the relative size of $\eta^2$ to $\beta$. However, notice that \eqref{gk:etalarge_1_ce} reduces to $\hat Q_{\hat x}\sim0$ for either choice and hence the pseudo-steady state is reached before solidification begins.

\subsubsection{Second time regime} 
As discussed previously, the choice of the time scale depends on the relative size of $\eta^2$ to $\beta$. Let us consider the distinguished limit $\beta\eta^{-2}=:\alpha=O(1)$ and choose $t=O(\alpha)$. From the previous time regime we have $Q=O(\eta^2)$ and $x,s,T=O(1)$. In the corresponding scaled variables $T=\tilde T$, $x=\tilde x$, $s=\tilde s$, $t=\alpha\tilde t$ and $Q=\eta^{2}\tilde Q$, the Stefan condition becomes
\begin{equation}
    {\tilde s}_{\tilde t}=-\tilde Q,\qquad \text{at }{\tilde x}={\tilde s},\label{gk:etalarge_2_Stefan}
\end{equation}
and, upon neglecting terms of order $\eta^{-2}$, the remaining equations take the form
\begin{subequations}
\begin{alignat}{3}
	&\gamma{\tilde Q}_{\tilde t}+\alpha\tilde Q=-{\tilde T}_{\tilde x\tilde t},\qquad &&0<\tilde x<\tilde s,\label{gk:etalarge_2_gk}\\
	&{\tilde Q}_{\tilde x}=0,\qquad &&0<\tilde x<\tilde s,\label{gk:etalarge_2_ce}\\
	&\tilde T = -1,\qquad &\text{at }&\tilde x=0,\\
	&\tilde T = 0,\qquad &\text{at }&\tilde x=\tilde s.
\end{alignat}
\end{subequations}
Taking derivatives with respect to $\tilde x$ in \eqref{gk:etalarge_2_gk} yields ${\tilde T}_{\tilde x\tilde x\tilde t}=0$. Additionally, from the previous time regime we have the matching condition ${\tilde T}_{\tilde x\tilde x}\sim0$ for $\tilde t=O(\beta^{-1})$, therefore ${\tilde T}_{\tilde x\tilde x}=0$ and hence, applying the boundary conditions, 
\begin{equation}
    \tilde T(\tilde x,\tilde t) = \frac{\tilde x}{\tilde s}-1,
\end{equation}
which reduces \eqref{gk:etalarge_2_gk} to 
\begin{equation}
    \gamma {\tilde Q}_{\tilde t}+\alpha\tilde Q=\frac{\tilde s_{\tilde t}}{\tilde s^2}.
\end{equation}
Using \eqref{gk:etalarge_2_Stefan} and the fact that ${\tilde Q}_{\tilde t}=(\partial \tilde{Q} / \partial \tilde{s}){\tilde s}_{\tilde t}$ we find that $\tilde Q$ is given by
\begin{equation}
    \tilde Q(\tilde s)=-\gamma^{-1}\left[\frac{1}{\tilde s}+\alpha(1-\tilde s)\right],
\end{equation}
where we have applied the matching condition $\tilde Q\sim-\gamma^{-1}$ and $\tilde s\sim1$ for $\tilde t=O(\beta^{-1})$. Replacing $\tilde Q$ in \eqref{gk:etalarge_2_Stefan} by this expression yields
\begin{equation}
    {\tilde s}_{\tilde t}=\gamma^{-1}\left[\frac{1}{\tilde s}+\alpha(1-\tilde s)\right].
\end{equation}
In particular, we observe that the initial solidification kinetics are similar to those predicted by the classical formulation, since $\tilde s\sim1$ for $\tilde t\ll1$ and hence $\tilde s_{\tilde t}\sim1/(\gamma\tilde s)$. In the original dimensionless variables we have $s_t=O(\eta^2)$, which indicates that non-local effects accelerate the solidification process. Interestingly, as the solid grows into the liquid, we find $\tilde s_{\tilde t}\to0$ as $\tilde s\to (1+\sqrt{1+4\alpha^{-1}})/2=\tilde s_0$ and therefore $\tilde Q\to0$ as the interface approaches this value. Hence, the problem becomes stationary at the end of this regime and the interface tends to the value $\tilde s_0=O(1)$. However, as the flux decreases and the time increases, the term $T_x$ enters the leading order problem and solidification starts again. The only possible balance for $\tilde Q\ll1$ and $\tilde t\gg1$ is when $\tilde Q=O(\eta^{-2})$ and $\tilde t=O(\eta^2)$, which also keeps the terms in the Stefan condition balanced, allowing solidification to start again as we enter the next regime.

\subsubsection{Third time regime} 
From the previous time regime we have $t=O(\beta)$ and $s,x,T,Q=O(1)$. Introducing the scaled variables $\bar t, \bar x,\bar s,\bar T,\bar Q$, the equations become
\begin{subequations}
\begin{alignat}{3}
	&\bar Q=-{\bar T}_{\bar x}-\alpha^{-1}{\bar T}_{\bar x\bar t},\qquad &&0<\bar x<\bar s,\label{gk:etalarge_3_gk}\\
	&{\bar Q}_{\bar x}=0,\qquad &&0<\bar x<\bar s,\label{gk:etalarge_3_ce}\\
	&\bar T = -1,\qquad &\text{at }&\bar x=0,\\
	&\bar T = 0,\qquad &\text{at }&\bar x=\bar s,\\
	&{\bar s}_{\bar t}=-\bar Q,\qquad &\text{at }&\bar x=\bar s.\label{gk:etalarge_3_Stefan}
\end{alignat}
\end{subequations}
Upon differentiating \eqref{gk:etalarge_3_gk} with respect to $\bar x$ gives $\alpha \bar T_{\bar x\bar x}+\bar T_{\bar x\bar x\bar t}=0$. Since the temperature profile $\bar T$ is linear in the previous time regime we find $\bar T_{\bar x\bar x}=0$ and hence
\begin{equation}
    \bar T(\bar x,\bar t)=\frac{\bar x}{\bar s}-1.
\end{equation}
Thus, the flux is given by
\begin{equation}
    \bar Q(\bar t)=-\frac{1}{\bar s}+\frac{\bar s_{\bar t}}{\alpha \bar s^2},
\end{equation}
and hence $\bar s$ is determined by
\begin{equation}\label{gk:etalarge_3_soln_s_ode}
    \bar s_{\bar t}=\frac{\alpha \bar s}{1+\alpha \bar s^2}.
\end{equation}
In particular, we find $\bar s_{\bar t}\sim\bar s^{-1}$ for $\bar t\gg1$, recovering so the classical solidification kinetics by the end of this time regime. The solution to \eqref{gk:etalarge_3_soln_s_ode} can be given implicitly,
\begin{equation}\label{gk:etalarge_3_soln_s_eq}
    \frac{1}{\alpha}\log(\bar s)+\frac{1}{2}\bar s^2=\bar t+\bar C
\end{equation}
or in terms of the LambertW function,
\begin{equation}
    \bar s(\bar t) = \sqrt{\frac{1}{\alpha}W\left(\alpha e^{2\alpha(\bar t+\bar C)}\right)}.
\end{equation}
The value of $C$ is found by imposing $\bar s\sim \tilde s_0$ for $\bar t=O(\eta^{-2})$ we find that
$\bar C=\tilde s_0^2/2 + \alpha^{-1} \log \tilde{s}_0$.

As suggested by the large-time behaviour of \eqref{gk:etalarge_3_soln_s_ode}, the classical dynamics are recovered in the fourth time regime, given by $\bar x,\bar s\sim\bar t^{1/2}$, $\bar Q\sim\bar t^{-1/2}$ and $\bar t=O(\eta^2)$.

\begin{figure}[h!]
    \centering
    \begin{subfigure}{.47\textwidth}
	\includegraphics[width=\textwidth]{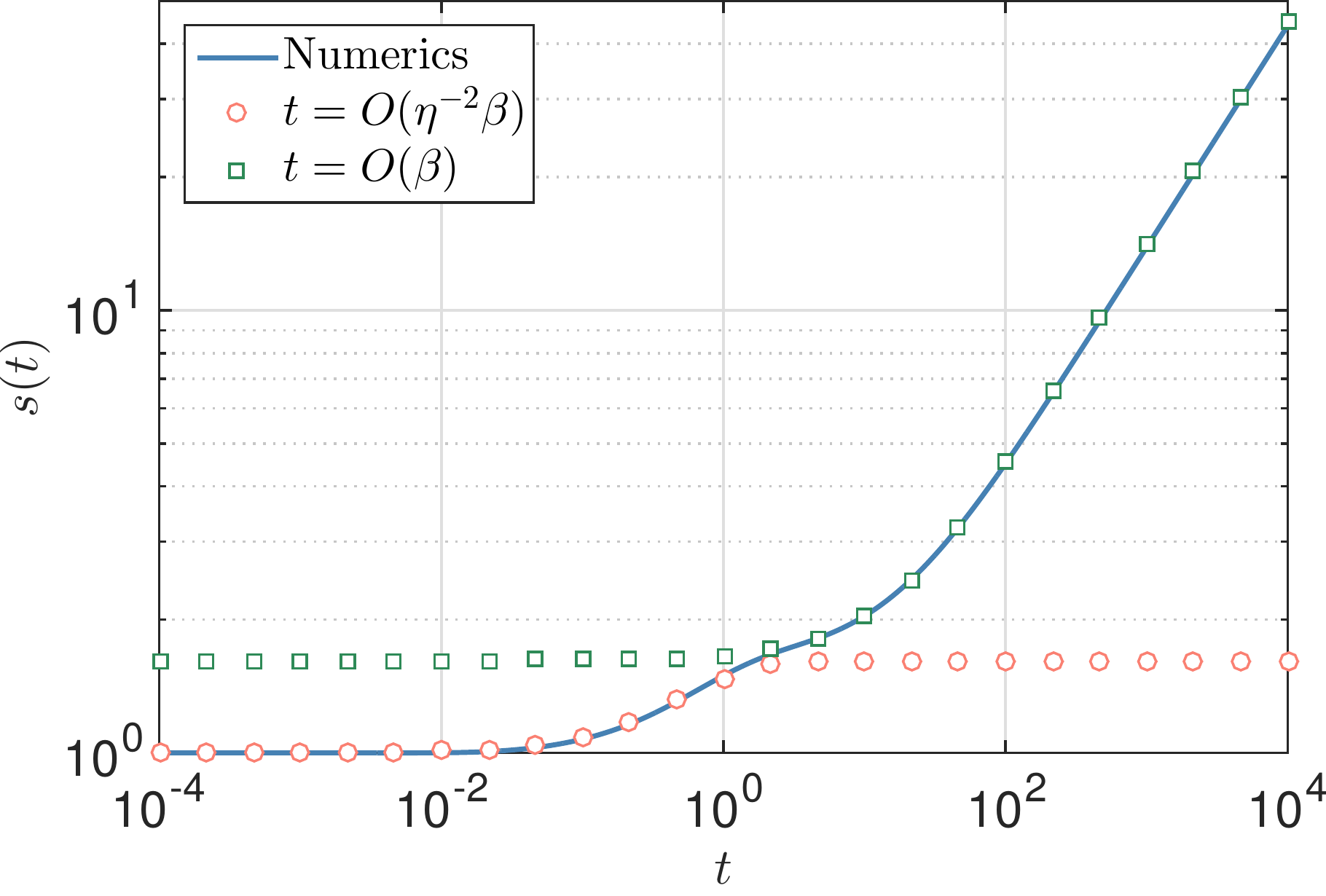}
    \caption{$\eta^2=10,\beta=10$.}
	\end{subfigure}
    ~
    \begin{subfigure}{.47\textwidth}
	\includegraphics[width=\textwidth]{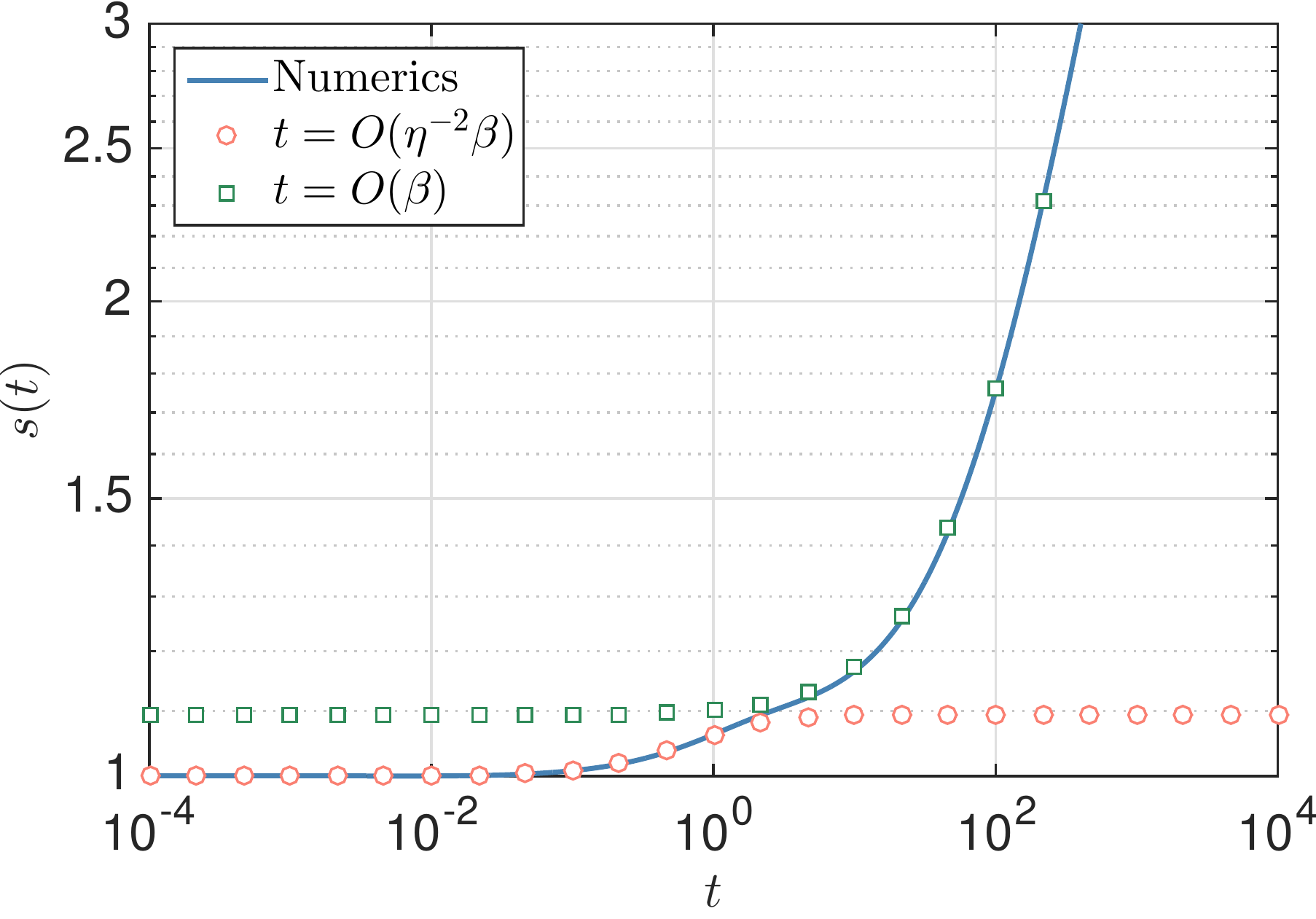}
    \caption{$\eta^2=10$, $\beta=100$.}
	\end{subfigure}
	
    \begin{subfigure}{.47\textwidth}
	\includegraphics[width=\textwidth]{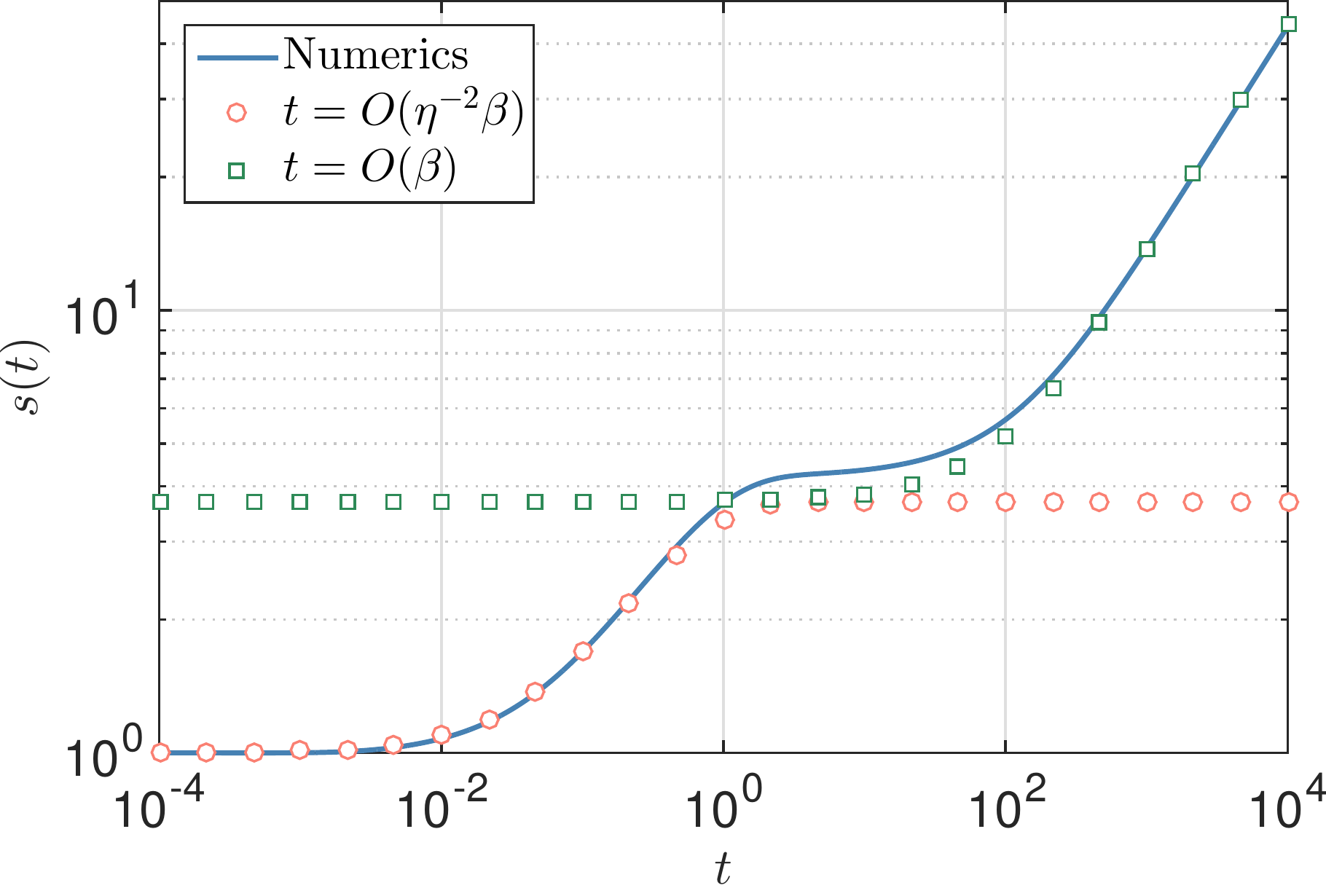}
    \caption{$\eta^2=100,\beta=10$.}
	\end{subfigure}
    ~
    \begin{subfigure}{.47\textwidth}
	\includegraphics[width=\textwidth]{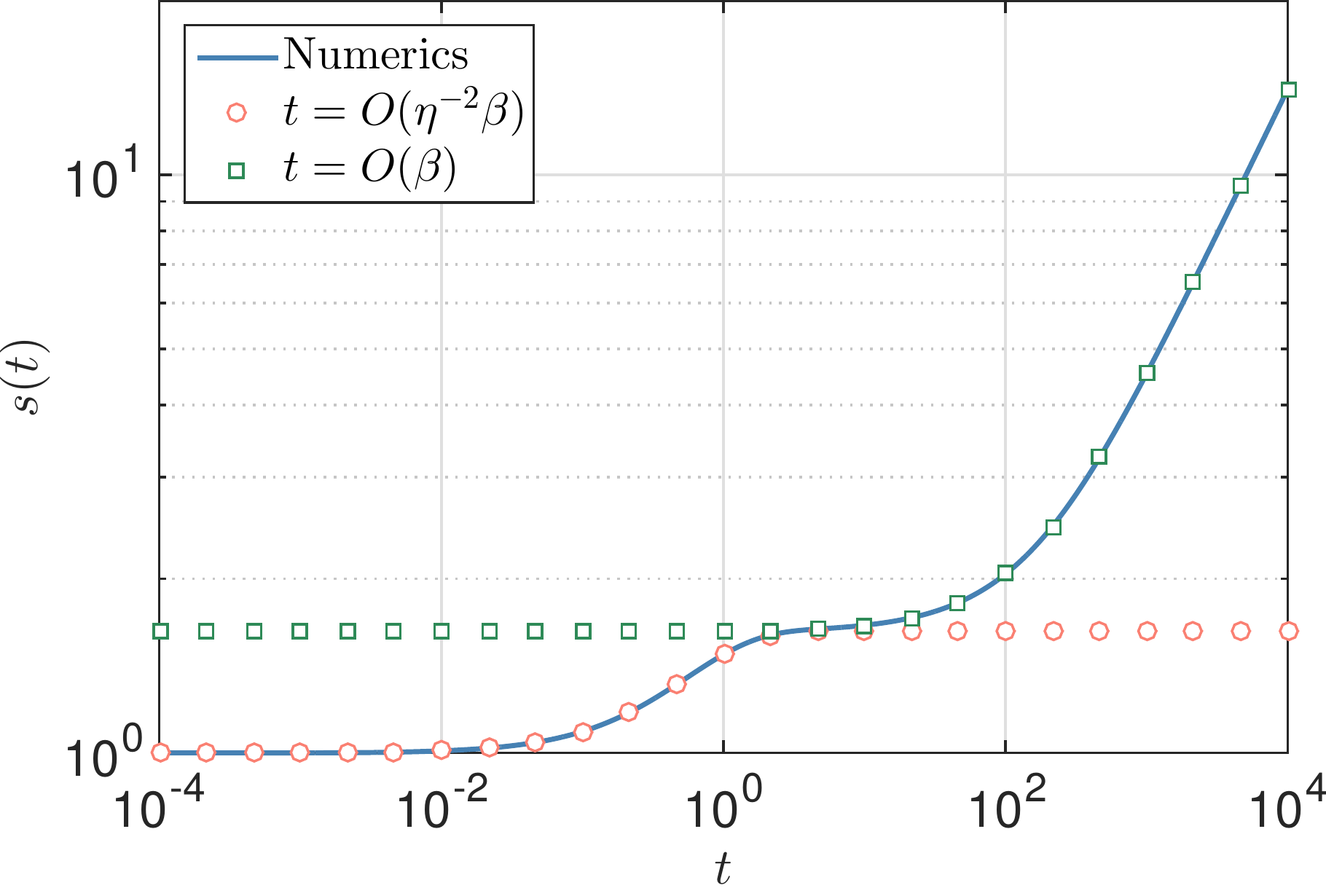}
    \caption{$\eta^2=100$, $\beta=100$.}
	\end{subfigure}
    \caption{Evolution of the interface for different values of $\beta$ and $\eta^2$, according to numerical simulation and the solutions of the second and third time regimes. The dimensionless mean free time has been set to unity.}
    \label{fig:num_asy2}
\end{figure}

\end{document}